\documentclass[iop,twocolumn,twocolappendix]{emulateapj}
\usepackage{amsmath}
\usepackage{hyperref}
\usepackage{epsfig}
\usepackage{color}
\begin{document}
\def\galaxiespprox{\mathrel{\vcenter{\offinterlineskip \hbox{$>$}
    \kern 0.3ex \hbox{$\sim$}}}}
\def\lapprox{\mathrel{\vcenter{\offinterlineskip \hbox{$<$}
    \kern 0.3ex \hbox{$\sim$}}}}
\newcommand{\beq}{\begin{equation}} 
\newcommand{\eeq}{\end{equation}}
\def\eps{\epsilon}
\def\epsw{\eps_{\rm w}}
\def\tH{t_{\rm H}}

\def\Edotw{\dot{E}_{\rm w}}
\def\Ewind{E_{\rm wind}}

\def\sigmastar{\sigma_{\ast}}
\def\vw{v_{\rm w}}
\def\vwten{v_{\rm w,10}}
\def\Sig15{\Sigma_{1.5}}

\def\kms{\rm\thinspace km~s^{-1}}
\def\keV{\rm\thinspace keV}
\def\cm{{\rm\thinspace cm}}
\def\pc{{\rm\thinspace pc}}
\def\kpc{{\rm\thinspace kpc}}
\def\pch{{\rm\thinspace pc} \thinspace h^{\rm -1}}
\def\Msun{\hbox{$\thinspace \rm M_{\odot}$}}
\def\Msunh{\hbox{$\thinspace  \it h^{\rm -1} \rm \thinspace M_{\odot} $}}
\def\ergs{\rm \thinspace erg \thinspace s^{-1}}
\def\yr{\rm yr^{-1}}
\def\Msunkpc{\hbox{$\thinspace \rm M_{\odot} \thinspace kpc^{-1.5}$}}

\def\pdot{\dot{p}}

\def\re{r_{\rm e}}
\def\rvir{r_{\rm vir}}

\def\Mvir{M_{\rm vir}}
\def\Mstar{M_{\ast}}
\def\Mbh{M_{\rm BH}}
\def\msigma{M_{\rm BH}-\sigmastar}
\def\lxsigma{L_X - \sigmastar}

\def\Mdotedd{\dot{M}_{\rm edd}}
\def\Mdotacc{\dot{M}_{\rm acc}}
\def\Mdotinf{\dot{M}_{\rm inf}}
\def\Mdotoutf{\dot{M}_{\rm outf}}
\def\Mdotbh{\dot{M}_{\rm BH}}
\def\mdotinf{\dot{m}_{\rm inf}}
\def\mdotoutf{\dot{m}_{\rm outf}}
\def\mdotedd{\dot{m}_{\rm Edd}}
\def\mdoteq{\dot{m}_{\rm eq}}
\def\mdotpl{\dot{m}_{\rm pl}}
\def\mdotacc{\dot{m}_{\rm acc}}
\def\mdotstar{\dot{m}_{\rm *,disk}}
\def\Mdotedd{\dot{M}_{\rm Edd}}
\def\Mdotpl{\dot{M}_{\rm pl}}
\def\Mdotbondi{\dot{M}_{\rm bondi}}

\def\strom{Str$\rm \ddot{o}$mgren$\thinspace$}

\shortauthors{Choi et al.}
\shorttitle{Physics of massive galaxy formation}

\title{Physics of Galactic Metals: Evolutionary Effects Due to Production, Distribution, Feedback \& Interaction with Black Holes}

\author{Ena~Choi\altaffilmark{1},
            Jeremiah~P.~Ostriker\altaffilmark{2,3},
            Thorsten~Naab\altaffilmark{4},
            Rachel~S.~Somerville\altaffilmark{1,5},
            Michaela~Hirschmann\altaffilmark{6},
            Alejandro~N\'u\~nez\altaffilmark{2},
            Chia-Yu~Hu\altaffilmark{4,5},
            Ludwig~Oser\altaffilmark{4}}

\affil{$^1$ Department of Physics and Astronomy, Rutgers, 
        The State University of New Jersey, NJ 08854, USA \\
        $^2$Department of Astronomy, Columbia University, New York, 
        NY 10027, USA \\
        $^3$Department of Astrophysical Sciences, Princeton University, Princeton, NJ 08544, USA\\
        $^4$Max-Planck-Institut f\"ur Astrophysik,
        Karl-Schwarzschild-Strasse 1, 85741 Garching, Germany \\
        $^5$Simons Center for Computational Astrophysics, New York, NY 10010, USA \\
        $^6$Sorbonne Universite's, UPMC-CNRS, UMR7095, Institut d' Astrophysique de Paris, F-75014 Paris, France\\
}

\begin{abstract}
We ask how the inclusion of various physical heating processes due to the metal 
content of gas affect the evolution of central massive galaxies and compute a suite of 
cosmological hydrodynamical simulations that follow these systems and their 
supermassive black holes. We use a smoothed particle hydrodynamics code with
a pressure-entropy formulation and a more accurate treatment of the metal 
production, turbulent diffusion and cooling rate based on individual element 
abundances. The feedback models include (1) AGN feedback via high velocity 
broad absorption line winds and Compton/photoionization heating, (2) explicit 
stellar feedback from multiple processes including powerful winds from supernova 
events, stellar winds from young massive stars and AGB stars as well as radiative 
heating within \strom spheres around massive stars, and (3) additional heating 
effects due to the presence of metals including grain photoelectric heating, 
metallicity dependent X-ray heating by nearby accreting black holes and from the 
cosmic X-ray background, which are the major improvements in our feedback 
model. With a suite of zoom-in simulations of 30 halos with 
$\Mvir \sim 10^{12.0-13.4}$, we show that energy and momentum budgeted from 
all feedback effects generate realistic galaxy properties. We explore the detailed 
role of each feedback model with three additional sets of simulations with varying 
input physics. We show that the metal induced heating mechanisms reduce the 
fraction of accreted stellar material by mainly suppressing the growth of diffuse 
small stellar systems at high redshift but overall have a relatively minor effect 
on the final stellar and gas properties of massive central galaxies. The inclusion of 
AGN feedback significantly improves the ability of our cosmological 
hydrodynamical simulations to yield realistic gas and stellar properties of massive 
galaxies with a reasonable fraction of the final stellar mass which is accreted from 
other galaxies.
\end{abstract}

\section{Introduction}\label{sec:intro}
The physical treatment of galaxy formation and evolution has steadily improved 
over recent years as the cosmological setting within which this process occurs 
becomes more definite and confirmed by multiple observations. Thus, the 
gravitational accumulation of mass on galactic scales from well-defined 
cosmological perturbations is well understood. Recent progress has focused on 
the ``feedback'', how the stars \citep[e.g.][]{2003MNRAS.339..289S,
2006MNRAS.373.1265O,2014MNRAS.445..581H,2015ApJ...804...18A} and 
black holes \citep[e.g.][]{2005ApJ...620L..79S, 2007MNRAS.380..877S,
2009MNRAS.398...53B,2012MNRAS.420.2662D} formed from the infalling gas 
can initiate processes which in turn alter that infall, drive outflows and change 
the rate of star formation from what would be guessed at from the simple picture 
of cosmological infall \citep[see][for a literature review]{2015ARA&amp;A..53...51S}. 
The two main drivers are supernovae (SN) and active galactic nuclei (AGN), but 
even normal star-formation of  massive and lower mass stars can have 
significant effects \citep{2013ApJ...770...25A,
2013MNRAS.436.1836R,2013MNRAS.428..129S,2014MNRAS.445..581H}.

The first generation of physics treatments for massive galaxies allowing for the 
incorporation of feedback concentrated on the mass, energy and momentum input 
by massive stars \citep[e.g.][]{2006MNRAS.373.1265O,2008MNRAS.387.1431D} 
and black holes \citep[e.g.][]{2010MNRAS.406L..55D,2010ApJ...722..642O,
2012ApJ...754..125C} found that the effects were dramatic. The ratio of baryons 
incorporated to those available was reduced by a factor of over ten, while the 
size of moderate mass galaxies and the duration of galaxy formation for these 
systems increased dramatically as kinetic feedback expelled gas. Some of this 
material returns at later times with increased angular momentum 
\citep{2014MNRAS.444.3357N,2015ApJ...804L..40G}. For those galaxies and for 
more massive ones the production of a circumgalactic medium due to outflow was 
found to have dramatic physical and observational consequences 
\citep[e.g.][]{2013MNRAS.432...89F,2015MNRAS.449..987F,
2016MNRAS.460.2157O}.

Recently the feedback effects of the chemical evolution have been  included, in 
addition to the mechanical effects. While most ``metals'' (everything heavier than 
H \& He) and particularly $\alpha$-process elements such as O and Mg are 
produced by young, massive stars, some elements, such as Fe are known to be 
produced primarily (but not exclusively) by older, lower mass exploding as type I 
SN and some elements, e.g. N, are significantly contributed by normal evolution of 
solar type stars. The line emission from these elements greatly increases cooling 
rates as compared to a (H,He) plasma, and recent papers such as 
\citet{2008MNRAS.385.2181F,2008MNRAS.387..577O,2009MNRAS.393...99W,
2009MNRAS.399..574W} or \citet{2013MNRAS.434.3142A} highlighted the 
dramatic consequences for the evolution of normal galaxies with the emphasis on 
moderate mass spiral systems. The net effect of the extra cooling processes is to 
again increase star formation rates somewhat, mitigating the mechanical effects 
of feedback. However, allowing only for the cooling effects of heavy elements is 
too one-sided and thus, not fully realistic.

There are also substantial heating effects due to the presence of metals 
\citep{draine2010physics}, and this paper attempts to spell out the most important 
ones. As we shall see, this reverses some of the cooling effects of metals recently 
included as we proceed, by successive approximations, towards a physically 
based and hopefully accurate treatment of galaxy formation. With successive 
iterations the situation becomes more complex, so, for example the metal 
component of interstellar gas is heated by X-rays both from the black holes 
within the galaxy and from the extra-galactic background, and photons from 
massive stars can heat gas via the photoelectric effect on metals distributed in the 
interstellar medium. All of these processes tend to increase the heating rate, 
causing a reduction of star formation, i.e. causing ``negative feedback''.

In this paper, we describe the implementation of the metal induced heating 
mechanisms as well as our more standard treatments of star-formation and other 
physical processes. We present in some detail the black hole-driven AGN 
feedback as it is different from that implemented in most codes in that it allows 
for broad absorption line winds, X-ray and UV output of both energy and 
momentum, rather than explicitly postulating ``radio'' feedback 
\citep[e.g.][]{2013MNRAS.436.3031V} or turning off cooling 
\citep[e.g.][]{2016arXiv160702151T}.  In addition, our SN feedback allows for three 
stages of SN remnant propagation (ejecta dominated, Sedov-Taylor, and
snowplow) with a relatively realistic treatment \citep[see][for details]{2017ApJ...836..204N}, without kinematically decoupling
the outflow particles nor disabling cooling for some fixed time interval as in many 
of SN feedback implementation in the literature. 

The purpose of this paper is to study the respective role of various feedback 
models implemented in our simulation on the physical properties of  central massive 
galaxies. In section~\ref{sec:z=0}, we present the 30 zoom-in simulations of halos 
with $\Mvir \sim 10^{12-13.4}$ with four different feedback models outlined in 
Section~\ref{sec:sims}, pointing out the changes induced by including the various 
physical processes. Finally, section~\ref{sec:dis} summarizes our conclusions.

\section{The simulations}\label{sec:sims}
The galaxy formation model and input physics adopted in the simulations have 
been improved from the previous papers. In this section we describe the new 
ingredients of the current model and we refer to the previous papers for the 
unchanged part. A comparison of present and past treatment is summarized in 
Table~\ref{tab:models}.

\begin{table*}
   \begin{center}
   \caption{Synoptic Table of Galaxy Model Development \label{tab:models}}
   \vskip+0.5truecm
    {
   \begin{tabular}{cc|c|ccccc}\hline\hline
   & \cite{2010ApJ...725.2312O,2012ApJ...744...63O}  & \cite{2015MNRAS.449.4105C} &\multicolumn{4}{c}{This Work} & \\
Input Physics & 
& 
ThSNnoMetal & 
Fiducial & 
NoAGN & 
NoZHeating & 
 \\ 
  \hline
{\bf Improved SPH}\tablenotemark{a} & No & Yes & Yes& Yes & Yes &   & \cr
{\bf Stellar feedback} & Thermal\tablenotemark{c} & Thermal\tablenotemark{c} & Mechanical\tablenotemark{d} & Mechanical\tablenotemark{d}  & Mechanical\tablenotemark{d} &  & \cr
{\bf AGN feedback} & No & Mechanical\tablenotemark{e} & Mechanical\tablenotemark{e} & No & Mechanical\tablenotemark{e} &  &  \cr
{\bf Metal production \& cooling} & No & No & Yes & Yes & Yes &  & \cr
{\bf Metal heating}\tablenotemark{b} & No & No & Yes & Yes & No & & \cr
  \hline\hline
   \end{tabular}}
   \end{center}
        \tablecomments{ A comparison of present and past treatment.
    \begin{flushleft}
    $^{a}$ Our improved SPH scheme includes density-independent pressure-entropy SPH formulation, 
    improved artificial viscosity and thermal conductivity, and a time-step limiter 
    for shock ambient particles \citep{2014MNRAS.443.1173H}.\\
    $^{b}$ Metal induced heating prescription includes the photoelectric heating and enhanced Compton heating by X-ray background. See Section~\ref{sec:Zheating} for details.\\
    $^{c}$ Thermal supernova feedback model from \cite{2003MNRAS.339..289S}.\\
        $^{d}$ Our mechanical stellar feedback model includes various processes such as ``snowplow'' SN winds, stellar winds 
    from young massive stars and AGB stars \citep{2017ApJ...836..204N}. See Section~\ref{sec:snfeedback} for details.\\
    $^{e}$ Our mechanical AGN feedback model includes high velocity 
    ($10,000 \kms$) broad absorption line winds  as well as photoionization 
    and Compton heating and associated radiation pressure from AGN \citep{2015MNRAS.449.4105C}. See Section~\ref{sec:agnfeedback} for details.
   \end{flushleft}}
\end{table*}

\subsection{Hydrodynamics code}
We use a modified version of the parallel smoothed particle hydrodynamics (SPH) 
code GADGET-3 \citep{2005MNRAS.364.1105S}. In order to avoid the numerical 
artifacts of the classical SPH code \citep[e.g.][]{2007MNRAS.380..963A} we use  
SPHGal \citep{2014MNRAS.443.1173H}, a modified version of GADGET-3 that 
includes a density-independent pressure-entropy SPH formulation  
\citep{2001MNRAS.323..743R,2013ApJ...768...44S,2013MNRAS.428.2840H}. To 
further improve over standard SPH, we adopt the Wendland $C^4$ kernel with 
200 neighboring particles \citep{2012MNRAS.425.1068D}, the improved artificial 
viscosity implementation \citep{2010MNRAS.408..669C}, and the artificial thermal 
conductivity following \cite{2012MNRAS.422.3037R}. Lastly, we employ a 
time-step limiter to ensure that neighboring particles have similar time-step, such 
that ambient particles do not remain inactive when a shock travels through 
\citep{2009ApJ...697L..99S,2012MNRAS.419..465D}. We refer the readers to 
\citet{2014MNRAS.443.1173H} for the performance of the new SPH schemes in 
the test problems.

\subsection{Star formation and chemical enrichment}
We adopted the star formation and chemical evolution model described in 
\cite{2013MNRAS.434.3142A}, which allows chemical enrichment by winds driven 
by Type~I Supernovae (SNe), Type~II SNe and asymptotic giant branch (AGB) 
stars with the chemical yields adopted from \citet{1999ApJS..125..439I,
1995ApJS..101..181W,2010MNRAS.403.1413K} respectively. We explicitly trace 
the mass in 11 different species, H, He, C, N, O, Ne, Mg, Si, S, Ca and Fe both for 
star and gas particles. Then the net cooling rate is calculated based on individual 
element abundances, temperature and density of gas (see 
\citet{2013MNRAS.434.3142A} for details). We adopted the cooling rate from 
\citet{2009MNRAS.393...99W} for optically thin gas in ionization equilibrium. We 
also include a redshift dependent UV/X-ray and cosmic microwave background with 
a modified \citet{2001cghr.confE..64H} spectrum.

We include a model for turbulent diffusion of gas-phase metals.
Following \citet{2013MNRAS.434.3142A}, we allow the metal enriched 
gas particles to mix their metals with neighboring gas particles using the 
standard SPH neighbor searches.

We stochastically form stars if the gas density exceeds a density threshold which 
we defined as $n_{\rm th} \equiv n_0 \left( T_{\rm gas}/ T_0 \right)^3 
\left( M_0 / M_{\rm gas}\right)^2 $ where critical threshold density and temperature 
are $n_0 = 2.0 \cm^{-3}$ and $T_0 = 12000$~K respectively and $M_0$ is the gas 
particle mass in fiducial resolution. We require that the gas density should be 
higher than the value for the Jeans gravitational instability of a mass 
$M_{\rm gas}$ at temperature $T_{\rm gas}$. The star formation rate is 
calculated as $d \rho_{\ast} /dt = \eta \rho_{\rm gas} /t_{\rm dyn}$ where 
$\rho_{\ast}$, $\rho_{\rm gas}$ and $t_{\rm dyn}$ are the stellar and gas densities, 
and local dynamical time for gas particle respectively. The star formation efficiency 
$\eta$ is set as $\eta=0.025$. In our model, star formation is feedback-regulated 
and star-forming regions grow in mass until sufficient new stars have formed to 
stop further collapse.

\subsection{Early stellar feedback and supernovae feedback}\label{sec:snfeedback}
We use the early stellar and SN feedback model adopted in \citet{2017ApJ...836..204N} (see 
section 2 in their paper for further details). We include the winds from young 
stars, UV heating from young stars \strom spheres, three-phase 
Supernova remnant input from both type I and type II SN feedback, and outflow 
and metals from dying low-mass AGB stars. Each of the processes is included as 
explicit physical processes communicating mass, metals, momentum and energy 
from stellar particles to surrounding gaseous particles.

In the early stellar feedback model, we include the effect of stellar winds as well 
as heating by the ionizing radiation from young massive stars before they explode 
as SNe. We distribute the momentum of winds from massive stars to the closest 
gas particles with the same amount of ejected mass and momentum as those of 
type II SN explosions evenly spread in time before the moment of SN explosion 
$t_{\rm SN}=$ 3 Myr. We also add the effects of ionizing radiation from massive 
stars. The cold gas with $T<10^4$~K within a \citet{1939ApJ....89..526S} radius 
from each star particle is gradually heated to $T=10^4$ K and is not allowed to 
cool below this temperature until it is no longer in an HII region, within the \strom 
radius \citep[see also][]{2012MNRAS.421.3522H}.

In the SN feedback model, a single SN event is assumed to eject mass in an 
outflow with a velocity $v_{\rm out,SN}=4,500 \kms$, a typical velocity of 
outflowing materials in SN. We distribute SN energy and momentum to the 
surrounding ISM from the SN event. We assume the SN outflows to be one of 
three characteristic SN remnant phases which determines how much of SN 
energy is transferred to gas particle as kinetic and thermal energy. Depending on 
the distance from the SN events, each neighboring gas particle is affected by one 
of the three successive phases: (i) momentum-conserving free expansion phase, 
(ii) energy-conserving Sedov-Taylor phase where SN energy is transferred with 
30\% as kinetic and 70\% as thermal, and (iii) the snowplow phase where 
radiative cooling becomes significant. In this model, the SN remnant initiates 
standard Sedov-Taylor blast-waves carrying energy as 30\% kinetic and 70\% 
thermal, and both amounts dissipate with distance from the SN, as described by 
the pressure-driven snowplow phase of SN remnants. See 
Appendix A in \citet{2017ApJ...836..204N} for a detailed description of the
implementation of SN feedback model. This model has 
provision for treating the interstellar medium (ISM) as multi-phase with most of 
the volume hot but most of the mass in warm or cool phases. The volume fraction 
in the hot phase is estimated by the volume weighted average temperature 
following the detailed high resolution simulations of \citet{2015ApJ...814....4L}.

Lastly, we include the feedback from the stars of low and intermediate initial mass 
via slow winds during an AGB phase. We transfer energy and momentum from 
old star particles to the neighboring gas particles in momentum-conserving way. 
The outflowing wind velocity of AGB stars is assumed to be 
$v_{\rm out, AGB}=10 \kms$, corresponding to typical outflowing velocities of AGB 
stars \citep[e.g.][]{1992A&amp;AS...93..121N}.

The metal-enriched  gas from dying stars is returned continuously to the ISM via 
winds from young stars, SNe and AGB stars. Over 30\% of total mass in stars is 
ultimately ejected via winds in our initial mass function assumption 
\citep{2001MNRAS.322..231K}, and able to induce the late star formation and 
quasi-stellar object (QSO) activity by feeding the central super massive black 
holes \citep[see also][]{2010ApJ...717..708C}.

\subsection{Black hole formation and growth}\label{sec:bh_acc}
In the simulations, the black holes are treated as collisionless sink particles and 
are seeded  in newly forming dark matter halos. The dark matter halos are 
identified on the fly during a simulation by  a friends-of-friends algorithm. The 
new black holes are seeded with mass of $10^5 \Msunh$ such that any halo 
above $1\times10^{11} \Msunh$ contains one black hole at its center if it does not 
already have any black hole. The dark matter halo threshold mass and black 
hole seed mass are set to roughly follow the \citet{1998AJ....115.2285M} relation 
and the theoretical calculations of \citet{2016arXiv160601909S}. The chosen 
seed mass makes a negligible contribution to the final black hole mass.

The black hole mass can grow via two channels: mergers with other black holes 
and accretion of gas. We allow the mergers between two black hole particles only 
when they fall within their local SPH smoothing lengths and their relative 
velocities are smaller than the local sound speed. As we cannot directly resolve 
the accretion disk of the black holes on sub-pc scales in the cosmological galaxy 
group-scale simulation, we estimate the rate of the gas infall onto the black 
hole with a Bondi-Hoyle-Lyttleton parameterization \citep{1939PCPS...34..405H,
1944MNRAS.104..273B,1952MNRAS.112..195B}. Following 
\cite{2005MNRAS.361..776S}, the gas accretion rate onto the central region 
around black hole is estimated as:
\begin{equation}
\dot{M}_{\rm{inf}}= 
\frac{4 \pi  G^{2} M_{\rm BH}^{2} \rho }
                            {(c_{\rm s}^2+ v^{2})^{3/2}},
\label{bondi_AA}
\end{equation}
where $\rho$, $c_{\rm s}$, and $v$ denote the density, the sound speed and the 
velocity of the gas relative to the black hole respectively. Several works adopting 
Bondi accretion in cosmological simulations often boost the accretion rate by a 
factor of $\alpha \sim 100$, \citep[][for a literature review]{2009MNRAS.398...53B}, 
here we do not employ additional ``boost'' factor $\alpha$ with regard to the 
accretion rate. 

We also include the soft Bondi criterion introduced in \cite{2012ApJ...754..125C} 
to avoid the unphysical accretion of unbound gas from outside the Bondi radius 
of the black hole. This criterion statistically limits the accretion to the gas within 
the Bondi radius. It also accounts for the size of the gas particle as the physical 
properties of each gas particle are smoothed within the kernel size in the 
{\it smoothed} particle hydrodynamics simulations. The full accretion is only 
allowed when the total volume of a gas particle is included within the Bondi 
radius. If a gas particle volume is partially included within the Bondi radius, its 
probability of being absorbed by the black hole is reduced. Finally, we include the 
free-fall timescale in the accretion rate in order to account for the time that it 
takes a particle to be accreted to black hole \citep{2012ApJ...754..125C}.

\subsection{Feedback from black holes}\label{sec:agnfeedback}
We use mechanical and radiative AGN feedback models described in 
\citet{2012ApJ...754..125C,2014MNRAS.442..440C}. Our AGN feedback model 
consists of two main components: (1) mechanical feedback as in the broad 
absorption line {\it winds}, which carry energy, mass and momentum into the 
surrounding gas. We also include (2) radiative feedback via the Compton and 
photoionization heating from the X-ray radiation from the accreting black hole, the 
radiation pressure associated with the heating, as well as the Eddington force. The 
emergent AGN spectrum and metal line heating are taken from 
\cite{2004MNRAS.347..144S}. In this section we summarize the main aspects of 
the AGN feedback model, for more details we refer the reader to 
\citet{2012ApJ...754..125C}.

\subsubsection{Mechanical AGN feedback}
In the presence of significant AGN winds, only a small fraction of the gas mass 
inflowing to the central region ultimately accretes to the black hole. AGN winds 
carry a mass with the outflowing rate given as,
\beq
\Mdotoutf= \Mdotinf - \Mdotacc, \label{eq:Mdot}
\eeq
where $\Mdotoutf$, $\Mdotinf$ and $\Mdotacc$ respectively denote the 
outflowing/inflowing mass rate and the mass accretion rate to the black hole. 
We assume a constant velocity for AGN wind $v_{\rm outf,AGN} =10,000$~$\kms$, 
corresponding to a typical broad absorption line wind velocity 
\citep[e.g.][]{2003ARA&amp;A..41..117C}.  The momentum flux carried by the wind 
is $\pdot = \Mdotoutf v_{\rm outf,AGN}$, and the kinetic energy rate carried by 
outflow will be given as,
\begin{subequations}
    \begin{eqnarray}
\Edotw & \equiv & \epsw \Mdotacc c^2, \label{eq:edotw1} \\
      &=& \frac{1}{2} \Mdotoutf v_{\rm outf,AGN}^2,\label{eq:edotw2} 
    \end{eqnarray}
\end{subequations}
where $\epsw$ denotes the feedback efficiency and we assume $\epsw=0.005$. 
The dimensionless quantity $\psi$ is defined as the ratio of the outflow rate to the 
accreted rate and given as,
\beq
\psi \equiv 2 \epsw c^2 / v_{\rm outf,AGN}^2=\Mdotoutf / \Mdotacc.
\label{eq:psi}
\eeq
Thus the black hole accretion rate depends on the mass inflow rate and the 
dimensionless parameter $\psi$ which is determined by mass and energy 
conservation as,
\beq
\Mdotacc = \Mdotinf \frac{1}{1+\psi}\label{eq:Mdot_sol}.
\eeq
We have $\psi \sim 9$ with our choice of the feedback efficiency $\epsw = 0.005$, 
and the wind velocity $v_{\rm outf,AGN} = 10,000$~$\kms$, thus bulk of mass 
entering the central region $\Mdotinf$ is expelled; 10\% 
($f_{\rm acc} = \Mdotacc/  \Mdotinf= 1/(1+\psi)$) of the inflowing mass is 
accreted onto the black hole and 90\% 
($f_{\rm outf} = \Mdotoutf/  \Mdotinf = \psi/(1+\psi)$) is expelled in an outflowing 
wind.

Among all gas particles entering the central region, the wind particles are 
stochastically selected with $f_{\rm outf}= \psi/(1+\psi)$. We deposit the wind 
mass and momentum by giving kicks to the selected wind particles. The 
direction of the wind is set to be parallel or anti-parallel to the direction of angular 
momentum of each gas particle, so that it results in a wind perpendicular to the 
disk plane \citep{2004ApJ...616..688P}, when the  black holes are surrounded by 
a rotating gas disk. The ejected wind particle shares its momentum and energy 
with two nearby gas particles and reproduce the shock heated momentum-driven 
flows. The residual energy increases the temperature of the impacted gas 
particles, therefore the total energy and momentum are conserved. This 
prescription gives a ratio of kinetic to thermal energy in the outflowing particles, 
similar to that in the standard Sedov-Taylor blast wave.

\subsubsection{Radiative AGN feedback}
We also include the heating via X-ray radiation from the accreting black hole. 
Strong X-ray radiation can be coupled to the surrounding gas according to an 
approximation described in \cite{2005MNRAS.358..168S}. The net luminosity 
flux from all black holes in the simulated zoom-in area is calculated at the 
position of each gas particle and the flux is converted to the net volume heating 
rate $\dot{E}$ by adopting the formulae that include  Compton and 
photoionization heating from \cite{2005MNRAS.358..168S}. Finally, the radiation 
pressure from the X-ray flux absorbed is included. Every fluid element absorbing 
energy $\Delta E$ from quasar radiation is given an extra momentum 
$\Delta p = \Delta E /c$ directed away from the black hole.

In many galaxy simulations with black holes, the gas accretion rate onto the black 
hole is limited to the Eddington rate. In our simulations, we do not limit the 
accretion rate onto the black holes, instead, we include the Eddington force acting 
on electrons in the neighboring gas through the hydrodynamic equations, directed 
radially away from black holes. In this way we allow that Super-Eddington gas 
accretion occasionally occurs so that the corresponding feedback effect naturally 
reduces the inflow and increases the outflow.

\subsubsection{Main aspects of the AGN feedback model}
The two AGN feedback mechanisms we have included originate from different 
parts of quasar's SED, i.e., mechanical feedback from UV radiation, and radiative 
feedback from X-ray radiation. Many previous high resolution hydrodynamical
simulations  \citep[e.g.][]{2000ApJ...543..686P} showed that the flux from strong 
UV radiation in the AGN spectra dominates for the momentum driven winds. The 
region where the winds are generated is very close to the accreting black holes, 
thus it is impossible to resolve this scale in cosmological simulations. Therefore 
we resort to the sub-grid modeling of ``mechanical feedback" via the broad
absorption line winds that have been accelerated by metal line trapping
\citep{2004ApJ...616..688P}. Meanwhile, the heating and associated pressure 
from AGN radiation are dominated by the moderately hard X-ray region 
($\sim 50$~keV) which is nearly independent of obscuration 
\citep{2004MNRAS.347..144S}. We therefore include this feedback effect 
following a standard atomic physics treatment. 

In \citet{2015MNRAS.449.4105C}, we found that AGN feedback has a very strong 
effect on the star formation and X-ray properties of galaxies and what matters 
most is the kind of feedback one includes. The mechanical AGN feedback via 
broad absorption line winds has more dramatic effects than the traditional thermal 
feedback treatment in many papers \citep[e.g.][]{2005Natur.433..604D,
2005MNRAS.361..776S,2009ApJ...690..802J,2015MNRAS.448.1504S}.
The two treatments - thermal vs. mechanical - put in the same total energy for a 
given accretion rate and given efficiency, but putting some fraction of the energy 
into mechanical rather than thermal increases the effectiveness in driving gas out 
of the galaxy as discussed in many recent papers 
\citep[e.g.][]{2012MNRAS.424..190G,2015ApJ...809...69S,2016MNRAS.461.1548B,
2017MNRAS.465.3291W}.

\citet{2014MNRAS.442..440C} showed that the mechanical and radiative AGN 
feedback can drive strong nuclear outflows as observed in many luminous 
quasars \citep[e.g.][]{2011ApJ...732....9G,2016MNRAS.459.3144Z} which can 
further progress on galaxy-wide scale removing significant fraction of gas from 
the host galaxies. Moreover, the relic AGN-driven winds from the height of recent 
activity found in the simulated galaxies provide a promising explanation for the 
moderate velocity outflows ($500 \-- 1000 \kms$) observed in post-starburst 
galaxies at intermediate redshift \cite[e.g.][]{2007ApJ...663L..77T} and in
quiescent nearby galaxies \citep[e.g.][]{2016Natur.533..504C}. The outflowing wind 
properties of our galaxy samples will be further discussed in a forthcoming paper 
(R. Brennan et al. 2016, in preparation).

\subsection{Metallicity dependent heating effect}\label{sec:Zheating}
\subsubsection{Photoelectric heating}
A dust grain can absorb an energetic far UV photon, excite it to higher energy 
and reemit as a ``photoelectron" \citep{1948ApJ...107....6S}. This photoelectric 
emission from dust grains dominates the heating of the diffuse ISM, and thus can 
have a significant effect on dynamical evolution of the ISM as well as on the 
subsequent star formation \citep{1978ApJS...36..595D,2001ApJS..134..263W}.
The precise modeling of photoelectric heating may require the estimation of the 
radial distance dependency from OB stars \citep[e.g.][]{2003ApJ...587..278W,
2011ApJ...730...11T}, self-shielding and shielding of radiation field 
\citep{2015MNRAS.454..238W,2016MNRAS.458.3528H}, as well as detailed 
model of dust physical properties such as grain sizes, compositions and charge 
state \citep{1994ApJ...427..822B}. However, it would be beyond the scope of this 
paper to fully account for all these factors which may affect the photoelectric 
heating rate. We therefore take the following simple prescription of the photoelectric
heating rate which is only dependent on the global star formation rate and the 
metallicity of gas which is related to the abundance of the smallest dust particles, 
which may primarily consist of polycyclic aromatic hydrocarbons.

From the numerical coefficient of photoelectric heating rate for the Galaxy from 
\citet{draine2010physics},  we added  temperature cutoff, 
$T_{\rm cut} = 10^4$~K so that only cool or 
cold gas is heated. Then, since the heating rate is proportional to the OB stars, i.e. 
the star formation rate, we added a factor for $(SFR/SFR_{MW})$. Then the 
factor $(SFR/SFR_{MW})$ is estimated by
\beq
(SFR/SFR_{MW})= \frac{Z}{Z_\odot} \frac{H(t)}{H_0}
\eeq
as the average value of this factor is just proportional to the metals produced
locally and inversely proportional to the time over which they were produced.
The photoelectric heating rate is thus estimated by
\beq
\Gamma = 1.4 \times 10^{-26} {\rm e}^{- T/10^4} \frac{Z}{Z_\odot} \frac{H(t)}{H_0} 
~~~{\rm erg/s~per~hydrogen~atom}.
\eeq
By construction it matches the local photoelectric heating rate given by 
\citet{draine2010physics} for the solar neighborhood.

\subsubsection{Cosmic X-ray background heating}
Energetic X-ray radiation from accreting black holes with a long mean-free path 
can make a significant contribution to the heating of the early universe, as 
it can easily escape the galaxies and impact the intergalactic medium at long 
range. We self-consistently include the heating effect from X-ray radiation of 
individual black holes within our simulation volume. However, as we use
zoom-in calculation with the limited volume, it is necessary to account for the 
effect of X-ray background, which may be able to pre-ionize large volumes of the 
intergalactic medium unreachable by UV sources. Thus, we include the effects of 
the low temperature gas heating by the cosmic X-ray background from accreting 
black holes in background galaxies as following.

First, we compute the cosmic X-ray background flux based on the information 
of the cosmic black hole mass accretion rates derived from the bolometric AGN
luminosity functions. The comoving number density of luminous quasars
reaches a maximum at $z_{\rm max} \sim 2$ \citep[e.g.][]{1995AJ....110...68S}.
We take the global black hole growth rates $\frac{d \rho_{\rm BH}}{dt}$ derived 
from the bolometric AGN luminosity functions of \citet{2009ApJ...690...20S} as:
\beq
\frac{d \rho_{\rm BH}(z)}{dt} = \frac{1 - \epsilon}{\epsilon c^2}
\int_0^{\infty} \Phi_L (L) L d {\rm log} L ,
\label{eq:drho}
\eeq
for example, 
$\frac{d \rho_{\rm BH}}{dt} (z=0)=3 \times 10^{-6} ~\Msun \rm yr^{-1} Mpc^{-3}$. 
Then, we take the cosmic evolution of black hole mass growth rates from 
\citet{2014ARA&amp;A..52..415M} and calculate the emissivity of AGN in a unit 
comoving volume of the universe at redshift $z$ as,
\beq
\varepsilon (z)= \epsilon \frac{d \rho_{\rm BH}(z)}{dt} c^2 ,
\eeq
where the radiative efficiency $\epsilon=0.1$. For example, 
$\varepsilon(z=0)=1.7 \times 10^{40}$ erg$\rm \thinspace s^{-1} Mpc^{-3}$.
The bolometric flux from the cumulative background AGN light at redshift $z$ is 
then
\beq
F_{\rm bk} (z) =  (1+z)^2 \frac{c}{4 \pi H_0} \int_z^{\infty} \frac{\varepsilon (z') dz'}{(1+z')[\Omega_M (1+z')^3 + \Omega_{\Lambda}]^{1/2}}.
\label{eq:fbk}
\eeq
The first $(1+z)^2$ term represents how physical area scales with redshift. We 
perform the integration of equation~\ref{eq:fbk} numerically and calculate the 
meta-galactic background flux level as a function of redshift. Then the net flux at 
the position of each gas particle is calculated as a sum of the global flux from the 
background AGN at given redshift and the flux from {\it all} black holes in the 
simulated zoom-in area based on their accretion rate.  All gas particles in the 
simulation are heated accordingly following heating rate described in 
\citet{2005MNRAS.358..168S}.

\subsubsection{Metallicity dependent X-ray heating}
For  cold gas exposed to the X-rays from an AGN the gas absorbs a certain 
number of X-ray photons which is proportional to the flux
\citep[see][for details]{2004MNRAS.347..144S}.
Most of this absorption is due to iron and similar heavy elements and the ejected 
electrons cause further ionizations and deposit much of the energy as heat. Thus 
the heating rate per H atom will be proportional to the X-ray flux as well as the 
metallicity. Thus in order to allow for metal line absorption, we include an 
additional metallicity dependent factor $(1+10 Z/Z_{\odot})$ in the X-ray heating 
rate by AGN radiation over and above Compton heating from 
\citet{2005MNRAS.358..168S} as,
\begin{equation}
S_1 = 4.1\times 10^{-35} (1.9\times 10^7 -T)\,\xi (1+10 Z/Z_{\odot}),
\end{equation}
where $\xi$ denotes the ionization parameter. Then the volume heating rate 
$\dot{E}$ by X-ray heating in cgs units is estimated as $\dot E = n^2 S_1$ where 
$n$ is the proton number density. For example, the radiative total X-ray heating 
for gas at solar metallicity is larger by a factor of $\sim 11$ than that of metal free 
gas.

\subsection{Cosmological `zoom-in' initial conditions}
The cosmological `zoom-in' initial conditions used in this study are described in 
detail in \citet{2010ApJ...725.2312O}. To achieve sufficiently high enough 
resolution to robustly model the evolution of galaxies, a sub-volume is extracted 
from a larger volume dark matter only simulation using a flat cosmology with 
parameters obtained from WMAP3 \citep[][$h=0.72, \;
\Omega_{\mathrm{b}}=0.044, \; \Omega_{\mathrm{dm}}=0.216, \;
\Omega_{\Lambda}=0.74, \; \sigma_8=0.77 $, and 
$\mathrm{n_s}=0.95$]{2007ApJS..170..377S}. At any given snapshot we trace 
back all particles close to the halos of interest from redshift zero. We replace
those particles with higher resolution gas and dark matter particles. Then new, 
high resolution initial conditions are re-simulated from redshift $z=43$ to $z=0$.

The simulations have been performed at two resolutions. (1) The reference resolution 
has the mass resolution for the star and gas particles are 
$m_{*,gas}=4.2 \times 10^{6} \Msunh$, and the dark matter particles have 
$m_{\mathrm{dm}} = 2.5 \times 10^{7} \Msunh$. We use the comoving 
gravitational softening lengths $\eps_{\mathrm{gas,star}} = 400 \rm \pch $ for the 
gas and star particles and $\eps_{\mathrm{halo}} = 890 \rm \pch$ for the dark 
matter which are scaled with the square root of the mass ratio following 
\citet{2001MNRAS.324..273D}. (2) The high resolution simulations have been 
performed with eight times better mass resolution than reference resolution, 
with $m_{*,gas}=5.3 \times 10^{5} \Msunh$, 
$m_{\mathrm{dm}} = 3.1 \times 10^{6} \Msunh$ and twice better spatial 
resolution with $\eps_{\mathrm{gas,star}} = 200 \rm \pch $ and 
$\eps_{\mathrm{halo}} = 450 \rm \pch$. In the Appendix, we discuss the resolution 
convergence of the galaxy physical quantities. 

The simulated halo masses range from  $1.4 \times 10^{12} \Msun$ 
$ \lesssim M_{\mathrm{vir}} \lesssim 2.3 \times 10^{13} \Msun$ at $z=0$ and the 
stellar mass of central galaxiex are $8.2 \times 10^{10} \Msun \lesssim M_{\ast} \lesssim 
1.5 \times 10^{12} \Msun$ at present day. These galaxies are well resolved with
$\approx 2.5 \times 10^4 - 4.8 \times 10^5$  stellar particles within the virial
radius ($r_{\mathrm{vir}} \equiv r_{200}$, the radius where the spherical 
over-density drops below 200 times the critical density of the universe at a given 
redshift).  

\begin{figure}
  \includegraphics[width=\columnwidth]{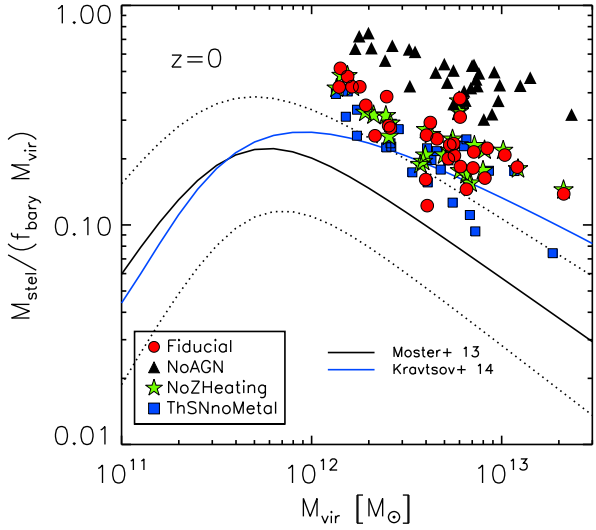}
  \caption{ Fraction of baryons converted to stars at $z=0$ relative to the universal 
  baryon fraction as a function of halo mass for central galaxies in different models:  
  fiducial model (red circles),  model without AGN feedback effect (NoAGN, black 
  triangles),  model without metal heating effect (NoZHeating, green stars), and 
  model with thermal   SN feedback and without metal enrichment
 (ThSNnoMetal, blue squares). The prediction of the fraction of stars inferred
 from observations via the abundance matching estimates  are shown in black and 
 blue solid lines from  \cite{2013MNRAS.428.3121M} and 
 \cite{2014arXiv1401.7329K}  respectively, with $1\sigma$ scatter region shown by 
 dotted lines. Compared to the fiducial model, the NoAGN model predicts 
 significantly higher conversion efficiencies especially at high halo masses while 
 the effect of metal induced heating mechanisms on the final stellar mass seems 
 negligible.}
    \label{fig:fbar}
\end{figure}

In order to study the effects of each of the physical modules we included, we 
investigate the galaxy properties by running the all 30 zoom-in simulations in 
reference resolution with four different models:

(1) {\bf Fiducial}: the reference model which includes all physical modules listed 
above, i.e., mechanical and radiative AGN feedback, stellar feedback with 
snowplow SN feedback, metal cooling and enrichment, and metal heating effect 
from photoelectric heating and cosmic X-ray background. Feedback related 
numerical parameters were calibrated against the the black hole mass -- stellar 
velocity dispersion ($\Mbh - \sigmastar$) relation and baryonic conversion rate
at $z=0$ (see Figure~\ref{fig:fbar} and \ref{fig:bh}).

(2) {\bf NoAGN}: without black hole and AGN feedback. This model isolates the 
effect of the AGN feedback.

(3) {\bf NoZHeating}: same as (1) fiducial but without the new ingredients of metal 
heating effects listed in Section~\ref{sec:Zheating} including metallicity dependent 
Compton heating, photoelectric heating and cosmic X-ray background heating.

(4) {\bf ThSNnoMetal}: from the previous paper \citep{2015MNRAS.449.4105C},
this model uses thermal SN feedback \citep{2003MNRAS.339..289S} instead
of ejective SN feedback described in Section~\ref{sec:snfeedback}. This model 
does not include metal enrichment, metal induced heating/cooling and early 
stellar feedback. Note that we presented 20 halos in 
\citet{2015MNRAS.449.4105C}, but we have performed 10 more zoom-in 
simulations for a fair comparison.

A comparative summary of the input physics for each model is given in 
Table~\ref{tab:models}. Throughout the paper, we focus on the 
central galaxies and discuss the respective role of various feedbacks
on the formation of massive galaxies.

\section{Variations of galaxy properties with physical models}\label{sec:z=0} 

In the following we will present simulated physical properties of the main four 
models and compare them to observations of various relations: stellar 
mass--halo mass relation, the relation between black hole mass and velocity 
dispersion, galaxies sizes, and stellar mass.

\subsection{The relation between stellar mass and halo mass}
Figure~\ref{fig:fbar} compares the simulated galaxies to the abundance 
matching results of \cite{2013MNRAS.428.3121M,2014arXiv1401.7329K}.
The mass of star particles within 10\% of the virial radius $r_{10}$ is defined as 
the stellar mass of simulated central galaxies. Note 
that we use the total baryonic and dark matter mass within the $\rvir$ for $\Mvir$
while the abundance matching models use the dark matter only simulations. As 
the baryonic physics can alter the dark matter distribution and reduce the halo 
masses, $\Mvir$ can be overestimated by 10~\% in the abundance matching 
models especially for the small halo mass range $\Mvir < 10^{10} \Msun$ 
\citep[e.g.][]{2013MNRAS.431.1366S, 2014MNRAS.440.2290M}. For our galaxy 
mass range, massive galaxies with $\Mvir > 10^{12} \Msun$, the stellar masses 
measured in the abundance matching model become increasingly sensitive to the 
aperture used to measure the stellar mass. Recently, \citet{2014arXiv1401.7329K} 
demonstrated that improved photometric techniques used to measure stellar 
mass lead to a significant effect on the stellar mass-to-halo mass relation 
predicted by abundance matching models. Their prediction is shown by the blue 
solid line. We see that the baryonic conversion efficiency in our models is very 
close to that inferred from abundance matching when AGN feedback is included. 
Models without AGN feedback over-predict the baryon abundances by a factor
of $2\--3$ at $z=0$.

\begin{figure}
 \epsfig{file=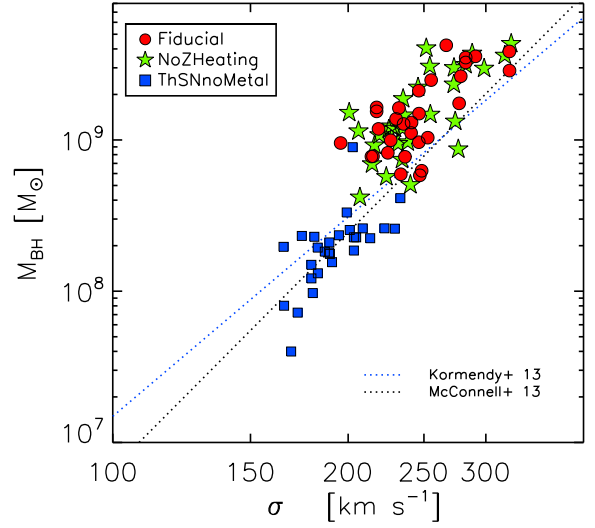,width=\columnwidth}
 \caption{The black hole mass -- stellar velocity dispersion ($\Mbh - \sigmastar$) 
 relation at $z=0$ of  fiducial model (red circles), model without metal heating 
 effect (NoZHeating, green stars), and model with thermal SN feedback and 
 without metal enrichment (ThSNnoMetal, blue squares).  The observed 
 M-$\sigmastar$ relations of the elliptical galaxies are shown in dotted lines: 
 blue dotted line is from \cite{2013ARA&amp;A..51..511K} and black dotted
 line from \cite{2013ApJ...764..184M}.
}\label{fig:bh}
\end{figure}

\subsection{The relation between black hole mass and stellar velocity dispersion}
Figure~\ref{fig:bh} shows the relation between the mass of central supermassive 
black holes and the velocity dispersion of galaxies and its comparison with 
observational data from \cite{2013ARA&amp;A..51..511K} and black dotted
line from \cite{2013ApJ...764..184M}.  We first determine the half-mass radii of 
stars within 10\% of the virial radius $r_{10}$ projected along the 20 randomly 
chosen directions. And the mean value is taken as a representative effective 
radius  $\re$ of each  galaxy. Then, the line-of-sight velocity dispersion 
$\sigmastar$ is calculated by considering all the stars within the half of the 
effective radius ($0.5 \thinspace \re$) along the three principal axes. Our new 
simulations show higher black hole masses as well as higher velocity dispersions 
compared to our previous work shown by the blue squares 
\citep{2015MNRAS.449.4105C}. This is because we have more gas available both 
for star formation and black hole accretion via enhanced gas metal-line cooling 
and recycled gas ejected from the death of stars in the old stellar population.

\subsection{Star formation rates}
\begin{figure}
 \epsfig{file=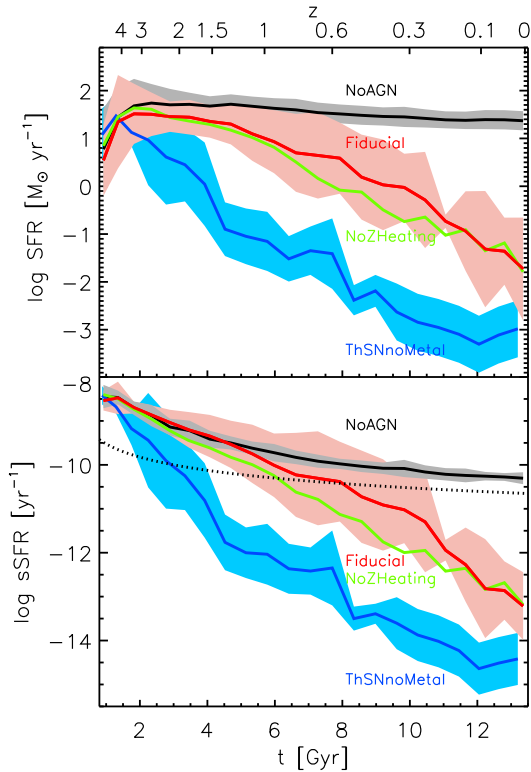,width=0.9\columnwidth}
 \caption{(Top) Averaged star formation rate over time for the 30 central galaxies 
 for different models:  fiducial model (red),  model without AGN  feedback effect 
 (NoAGN, black),  model without metal heating effect  (NoZHeating, green), and 
 model with thermal SN feedback and without metal  enrichment (ThSNnoMetal, 
 blue). The solid lines show the average value and  shaded regions illustrate the 
 $1\sigma$  scatter. For clarity of display, the  $1\sigma$ region of the 
 NoZHeating model is not shown. (Bottom) same as in the  top panel but for the 
 median specific star formation rates. The dotted black line  indicates the specific 
 star formation rates equal to~$ 0.3/\tH$, commonly used criteria separating 
 quiescent and star forming galaxies  \citep[e.g.][]{2008ApJ...688..770F}. The 
 NoAGN feedback model (black) stays above this criteria, constantly star 
 forming through out the evolution.}
    \label{fig:sfr}
\end{figure}

In Figure~\ref{fig:sfr}, we address ``quenching'' of star formation and show the 
dramatic effects that the various physical processes have on the mean star 
formation rate within 10\% of the virial radius $r_{10}$ (top panel) and the specific 
star formation rate (bottom panel) of 30 central galaxies in our four models as a 
function of redshift.  The star formation rate of all models peaks at $z \sim 3$ and 
drops rapidly afterwards, except for the NoAGN model. With mechanical and 
radiative AGN feedback included, the star formation is very rapidly quenched by 
the effective removal of gas via AGN-driven large-scale winds 
(see \citet{2016arXiv161103869P} for detailed discussion on the 
quenching timescale of the fiducial galaxies).

Compared to NoZHeating model, metal-dependent heating in the fiducial run has 
the effect of reducing the star formation rate around the redshift of quenching at 
$2<z<4$. This is because both photoelectric heating and cosmic X-ray 
background heating from accreting black holes mainly affect diffuse and small 
stellar system effectively suppressing their growth.

With metal induced heating, the formation of low-mass stellar systems is efficiently 
delayed, but the heated gas will later cool down and come back to the central 
system, therefore the star formation rate later compensates at $z<1$.

In the bottom panel, we show the median specific star formation rates of four 
models.  We also show the commonly used criteria separating quiescent 
and star forming galaxies \citep{2008ApJ...688..770F}, the specific star 
formation rates equal to~$ 0.3/\tH$ in dotted black line. The NoAGN feedback 
model (black) stays above this criteria constantly, star forming throughout the 
simulation. By this criterion the fiducial models cross over to the red sequence 
and become quiescent between redshift $z=1.5$ and  $z=0.3$. Again 
compared to our previous models 
\citep[ThSNnoMetal,][]{2015MNRAS.449.4105C}, allowing for the additional 
metal cooling from recycled gas increases star formation rates (red curves) but 
leaves them at $z=0$ still more than two orders of magnitude below the NoAGN 
models.

\subsection{Galaxy sizes and velocity dispersions}
\begin{figure}
 \epsfig{file=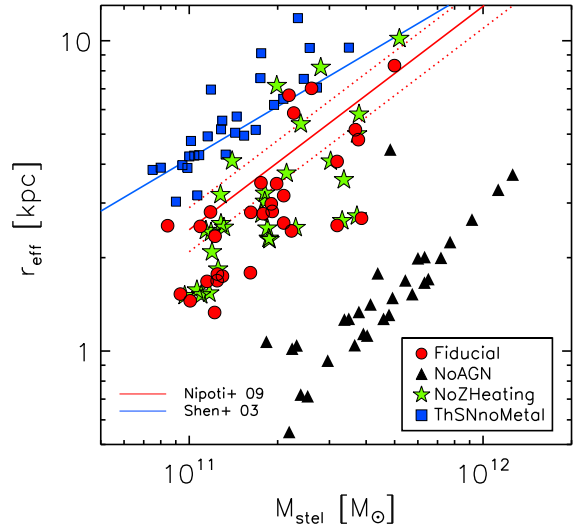,width=\columnwidth}
 \caption{Projected stellar half-mass radii of the simulated galaxies versus stellar 
 masses at redshifts z=0 for four models:  fiducial model (red),  model without 
 AGN feedback effect (NoAGN, black),  model without metal heating effect 
 (NoZHeating, green), and  model with thermal SN feedback and without metal 
 enrichment (ThSNnoMetal, blue). The red solid line indicates the observed 
 size-mass relation of the SLACS sample of local early-type galaxies 
 \cite{2009ApJ...706L..86N}, with the $1\sigma$ scatter given by the dotted 
 lines. The blue solid line is from \cite{2003MNRAS.343..978S}. } 
    \label{fig:size}
\end{figure}

\begin{figure}
\epsfig{file=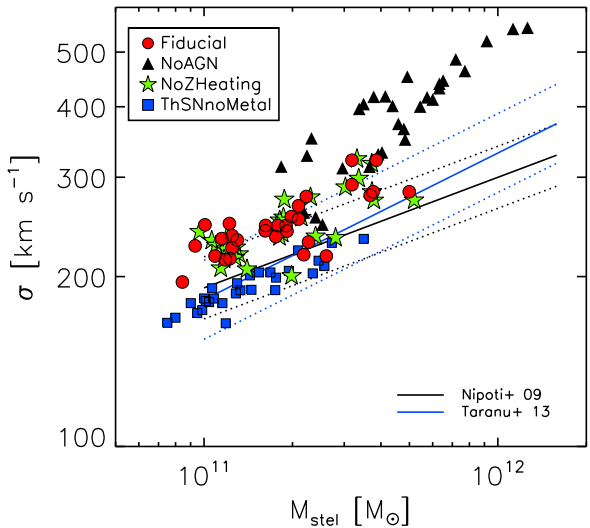,width=\columnwidth}
 \caption{The projected velocity dispersion measured within 0.5 $\re$ as a 
 function of stellar masses at z=0 for central galaxies in different models: 
 fiducial model (red),  model without AGN feedback effect (NoAGN, black),  model 
 without metal heating effect (NoZHeating, green), and  model with thermal
 SN feedback and without metal enrichment (ThSNnoMetal, blue). The 
 observational relations for local early type galaxies respectively from 
 \citet{2009ApJ...706L..86N} and \citet{2013ApJ...778...61T} are shown by the 
 black and blue solid lines with the dotted lines indicating the $1\sigma$ scatter.}
    \label{fig:sigma}
\end{figure}

\begin{figure*}
  \begin{minipage}{\textwidth}
    \centering
    \includegraphics[width=.45\textwidth]{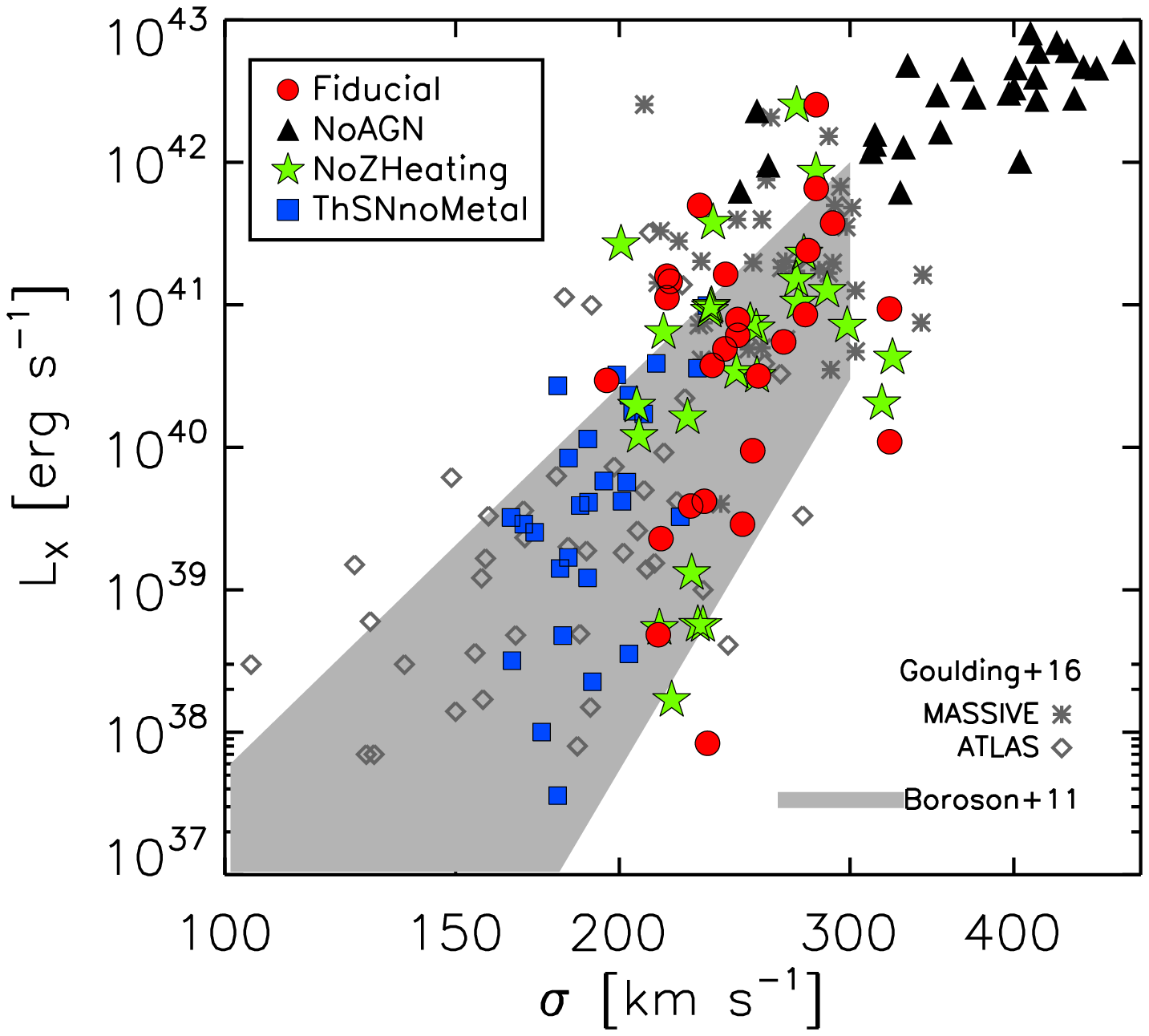}\quad
    \includegraphics[width=.45\textwidth]{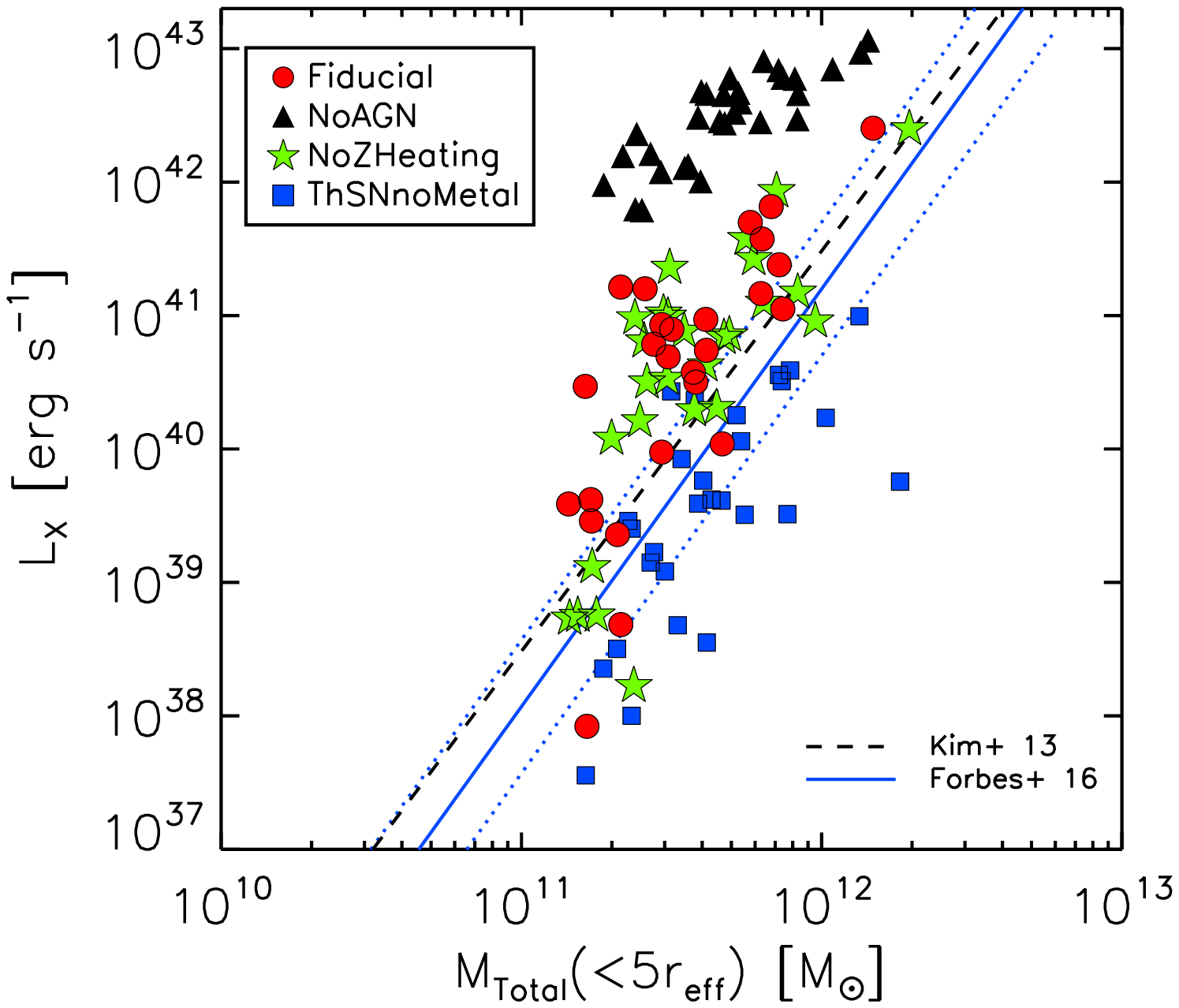}
    \caption{(left) $L_X$-$\sigmastar$ relation and (right) 
    $L_X$-$M_{\rm total}(<5 \re)$ relation at z=0     four models:  fiducial model 
    (red),  model without AGN feedback effect (NoAGN, black),  model without 
    metal heating effect (NoZHeating, green), and  model with thermal SN 
    feedback and without metal enrichment (ThSNnoMetal, blue). Observed 
    relations and data for the normal early-type galaxies are from 
    \citet{2011ApJ...729...12B} and \citet{2016ApJ...826..167G} (left) and
    \citet{2013ApJ...776..116K} and \citet{2017MNRAS.464L..26F} (right).}
    \label{fig:lx}
  \end{minipage}\\[1em]
\end{figure*}

In Figure~\ref{fig:size}, we show the projected half mass radius $\re$, i.e., the 
radius that encloses 50\% of the stellar mass in projection, as a function of the 
galaxy mass at $z=0$. As described above, we calculate the half-mass radii of 
stars within $r_{10}$ projected along the 20 randomly chosen directions of the 
main stellar body, and the mean value is taken as a representative effective 
radius $\re$ of each galaxy. The sizes of the fiducial galaxies at high stellar 
masses simulated with AGN feedback are in good agreement with observations 
of early-type galaxies from \cite{2009ApJ...706L..86N}, but we tend to produce 
typically too small sizes for lower mass galaxies.

The sizes of galaxies in NoAGN feedback models are even more smaller; their 
effective radii are $\sim 5$ times smaller compared to observed ones at given 
stellar mass. The continuous star formation without AGN feedback in the central 
regions of galaxies leads to a concentrated stellar mass profile 
\citep{2012MNRAS.422.3081M}. Moreover, the galaxies are less puffed up by 
minor mergers as the absence of AGN feedback decreases the fraction of 
accreted material \citep{2013MNRAS.433.3297D,2016MNRAS.463.3948D}.
In the absence of AGN feedback, in situ star formation dominates over 
accreted star formation at all times \citep{2012MNRAS.425..641L,
2013MNRAS.436.2929H}, therefore relatively fewer stars are added to the 
outskirts of galaxies.

The inclusion of physical heating processes due to the metal content of gas 
slightly decreases the galaxy sizes. The metal-dependent heating mainly affects 
the smaller systems decreasing the total mass of accreted stars. Therefore, the 
accreted star fraction decreases when we include metal-dependent heating and 
this leads to smaller galaxy sizes.

Similarly in the ThSNnoMetal model, even more small stellar systems are 
accreted to the central galaxy due to insufficient stellar feedback to suppress star
formation in the small building blocks, leading to a significant size growth of 
galaxies at late times.  In addition, the lack of metal cooling in this model prevents 
gas from concentrating in the central region of galaxy, thus contributing to the 
size increase.

From \citet{2012ApJ...744...63O}, our galaxy evolution code has been improved 
with addition of the density independent SPH formulation, the metal enrichment, 
and the subsequent metal-line cooling. All of these changes enhance the gas 
cooling and star formation especially for the central galaxies. In addition, the SN 
feedback model is changed from thermal to mechanical, which drives more 
powerful winds. However, this change also challenges the late time-star formation 
quenching in central galaxies as the material removed from lower mass 
progenitors can be accreted to more massive ones at later times 
\cite[e.g.][]{2008MNRAS.387..577O,2013MNRAS.436.2929H}. The previous 
version of our physical model was able to reproduce the quiescent galaxies
via gravitational heating \citep[e.g.][]{2009ApJ...697L..38J} and sufficient size 
growth afterwards even in the absence of AGN feedback, but this seems to be 
due to some missing physics. The gravitational heating helps but by itself is 
insufficient to sufficiently quench extended star formation in massive galaxies. 
Overall, appropriately predicting the final effective radius and its evolution provides 
a strong test of the accuracy of the physical modeling. A further study of the 
physical mechanisms determining the size evolution of galaxies will be 
discussed in a forthcoming paper \citep{choi16_size}.

Figure~\ref{fig:sigma} shows the Faber-Jackson relation 
\citep{1976ApJ...204..668F}, the relation between the stellar velocity dispersion 
$\sigmastar$ and stellar mass for simulated galaxies at $z=0$. The line-of-sight 
velocity dispersions $\sigmastar$ are measured for all stars within $0.5 \times \re$ 
along the three principal axes. The observed relations with $1 \sigma$ ranges 
are from \cite{2009ApJ...706L..86N} based on SLACS sample of local early-type 
galaxies at $z=0$ and from \cite{2013ApJ...778...61T} based on SDSS catalog 
respectively. All simulations except the NoAGN feedback model are in good 
agreement with the data, but the fiducial model tends to have overall higher 
velocity dispersions. Compared to the ThSNnoMetal model, the fiducial model 
shows higher velocity dispersion due to the increased central star formation 
driven by enhanced gas metal-line cooling and recycled gas ejected from the 
old stellar population. The NoAGN model significantly over-predicts the velocity 
dispersion of normal massive elliptical galaxies due to the enhanced star formation 
in the central region of galaxies.

\begin{figure}
\epsfig{file=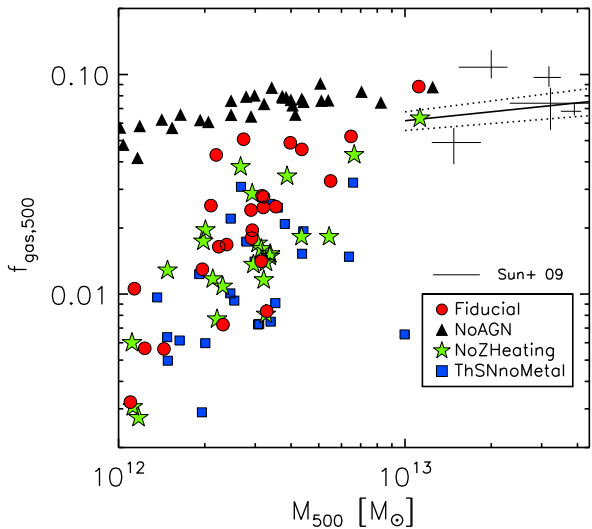,width=\columnwidth}
 \caption{The enclosed gas mass fraction within $r_{500}$,  
 $f_{\rm gas,500} - M_{500}$ relation for four models:  fiducial model (red), model  
 without AGN feedback effect (NoAGN, black),  model without metal heating 
 effect (NoZHeating, green), and  model with thermal SN feedback and without 
 metal enrichment (ThSNnoMetal, blue). NoZHeating model shows lower gas 
 fraction compared to the fiducial model. The black crosses and black solid lines 
 are the observed data and fitted $f_{\rm gas,500} - M_{500}$ relation of the 
 nearby galaxy groups with $M_{500}=10^{13-14} \Msun$ from 
 \citet{2009ApJ...693.1142S}.}
\label{fig:fgas}
\end{figure}

\subsection{X-ray luminosity and gas mass fraction}
Recent X-ray observations have shown that hot gas halos are ubiquitous from 
galaxy clusters, groups and galaxies with mass down to 
$\Mstar \sim 10^{11} \Msun$ \citep[e.g.][]{2015MNRAS.449.3806A}. We explore 
the impact of various feedback models on hot gas in the simulated galaxy halos. 
In Figure~\ref{fig:lx}, we show the X-ray luminosity of the hot gas versus the 
stellar velocity dispersion $\sigmastar$ and total mass within $5 \re$ of the 
simulated central galaxies at $z=0$ for four models. Following 
\cite{1995ApJ...451..436C}, we calculate the X-ray luminosity for the simulated 
galaxies in {\it Chandra} bands (0.3-8 keV) including 
bremsstrahlung radiation and metal-line emission from all relevant species 
measured. For ThSNnoMetal model, we assume solar abundance.
For $L_X - \sigma$, we show the observed relation from 
\citet{2011ApJ...729...12B} and data from \citet{2016ApJ...826..167G}
who archived {\it Chandra} X-ray observation of local elliptical galaxies and 
combined the data from MASSIVE galaxy survey \citep{2014ApJ...795..158M} 
and ATLAS$^{\rm 3D}$ survey. For $L_X \-- M_{\rm Total}(<5r_{\rm eff})$, we 
compare our simulated galaxy properties to \citet{2013ApJ...776..116K}
and \citet{2017MNRAS.464L..26F} who recently reexamined the dynamical
masses of the galaxies with the globular cluster kinematics.

In the fiducial model, the mechanical energy output of the black hole at the peak 
epoch of galaxy formation drives galactic outflows and expels large amounts of hot 
gas from the galactic potential. This large scale winds result in overall lower gas 
density and thus significantly reduce the gas X-ray luminosity as 
$L_X \propto n^2$. The fiducial model with mechanical AGN feedback 
reproduces X-ray luminosity within the observed range as well as the 
observed scatter with the two orders of magnitude spread in $L_X$. Conversely, 
the NoAGN model shows much higher X-ray luminosity up to $\sim$ 1-2 orders 
of magnitude higher compared to the observations elliptical galaxies.  In our 
simulations, X-ray luminosity of hot halos of massive galaxies is most
sensitive to AGN feedback consistent with many previous works including 
\citet{2010MNRAS.406..822M,2014MNRAS.441.1270L,2015MNRAS.451.3868L,
2016MNRAS.tmp.1592L}. In particular, \citet{2014MNRAS.441.1270L} showed 
that their AGN feedback model with the heating temperature of 
$\Delta T_{\rm heat} = 10^8 $ K produced the observed X-ray luminosity - mass 
relation in agreement with observations, and also noted that X-ray luminosities
of simulated galaxies are sensitive to the choice of the heating temperature
parameter in their AGN feedback model.

In Figure~\ref{fig:fgas} we show the gas mass fraction of all simulated halos and 
compare them to observations. We measure the fraction of gas mass to the total 
mass within $r_{500}$, the radius where the spherical over-density drops below 
500 times the critical density of the universe at $z=0$. The observed 
$f_{\rm gas,500}-M_{500}$ relation derived from 43 nearby galaxy groups with 
$ M_{500} = 10^{13-14} \Msun$ from \citet{2009ApJ...693.1142S} is shown in 
black solid line with 1$\sigma$ scatter in dotted lines. We also overplot the 
observed  gas mass fraction of galaxy groups from the same paper in black 
crosses. In the fiducial model, the AGN-driven strong wind results in much lower 
gas mass fraction compared to the NoAGN model especially for the low mass 
galaxies because a lot of hot gas is swept up and removed from the halo by AGN 
driven galactic outflows. The inclusion of metal induced heating also lowers the 
gas fraction, but its effect is relatively small. 

\begin{figure}
\epsfig{file=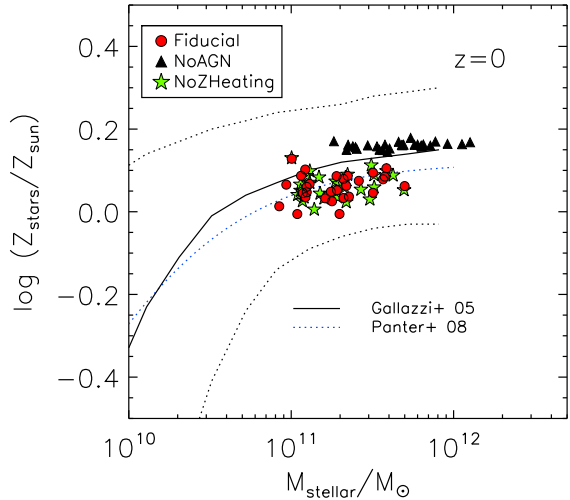,width=\columnwidth}
\caption{The mass metallicity relation - stellar metallicity versus galaxy stellar 
mass at $z=0$ for three models:  fiducial model (red circles),  model without 
AGN feedback effect (NoAGN, black triagnles), and  model without metal heating 
effect (NoZHeating, green stars). ThSNnoMetal model is not shown as this
model assumed a primordial metal abundance and did not include metal 
enrichment. Blue dotted line and black solid line indicate the local 
mass-metallicity relation for the SDSS galaxy population from
\citet{2008MNRAS.391.1117P} and from \citet{2005MNRAS.362...41G} 
respectively. Fiducial model is in good agreement with observations. }
\label{fig:mzr}
\end{figure}

\subsection{Mass-Metallicity Relation}
The relations between stellar metallicity and mass of simulated galaxies are 
shown in Figure~\ref{fig:mzr} and compared with observations from SDSS from 
\cite{2008MNRAS.391.1117P} and \cite{2005MNRAS.362...41G}. The NoAGN 
feedback shows systemically higher metallicity by $\sim 0.05$ dex compared to 
the fiducial model simulated with AGN feedback.  The higher metallicity in the 
NoAGN feedback model driven by the excessive late-time star formation out of 
metal enriched gas in absence of effective star formation quenching mechanism. 
A more detailed investigation of the impact of AGN feedback of individual metal 
abundances and metallicity gradients of stellar populations will be given in 
Hirschmann et al. {\it in preparation}.

\subsection{The Two Phases of Galaxy Formation}   
\begin{figure*}
  \begin{minipage}{\textwidth}
    \centering
    \includegraphics[width=.4\textwidth]{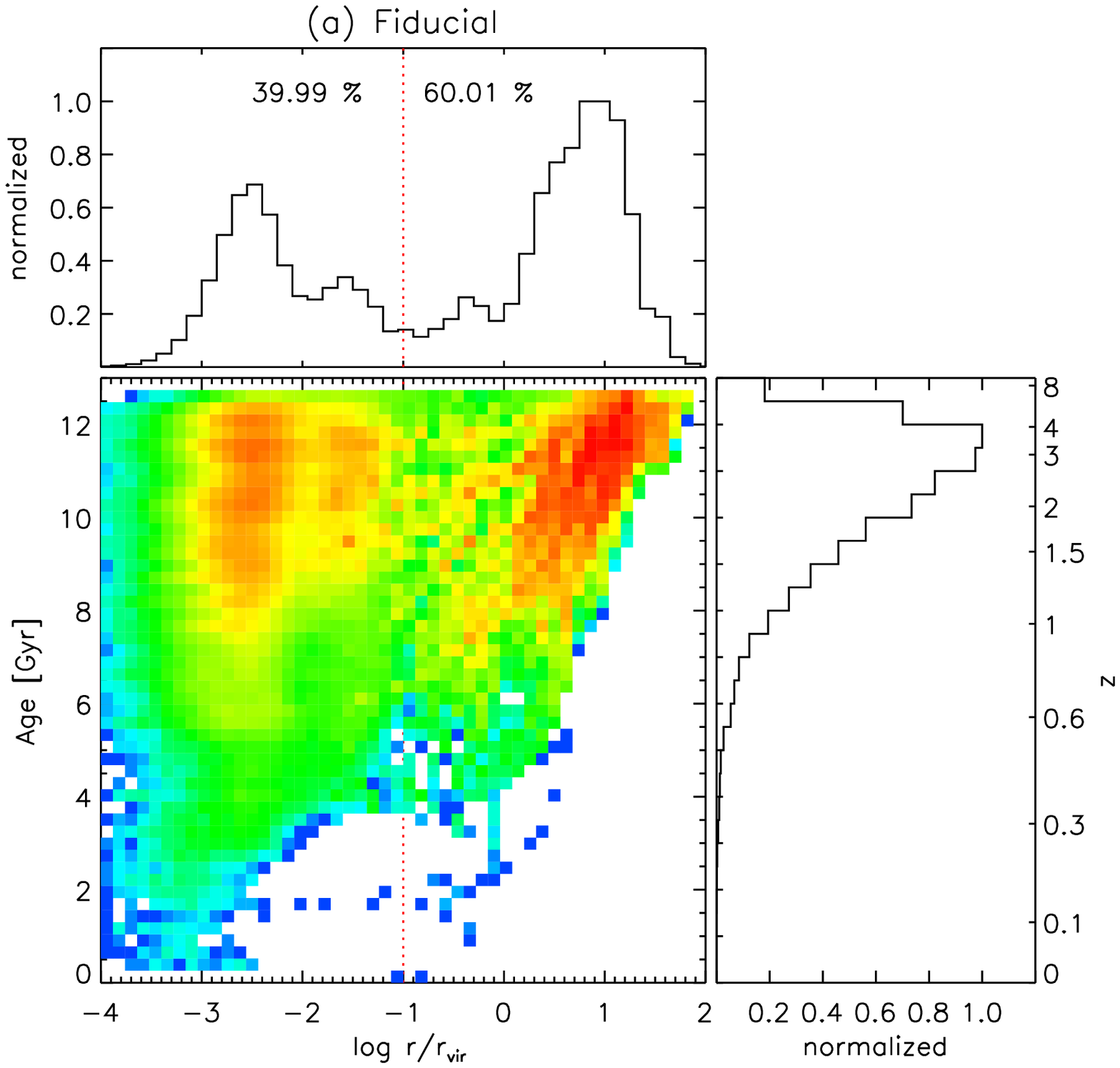}\quad
    \includegraphics[width=.4\textwidth]{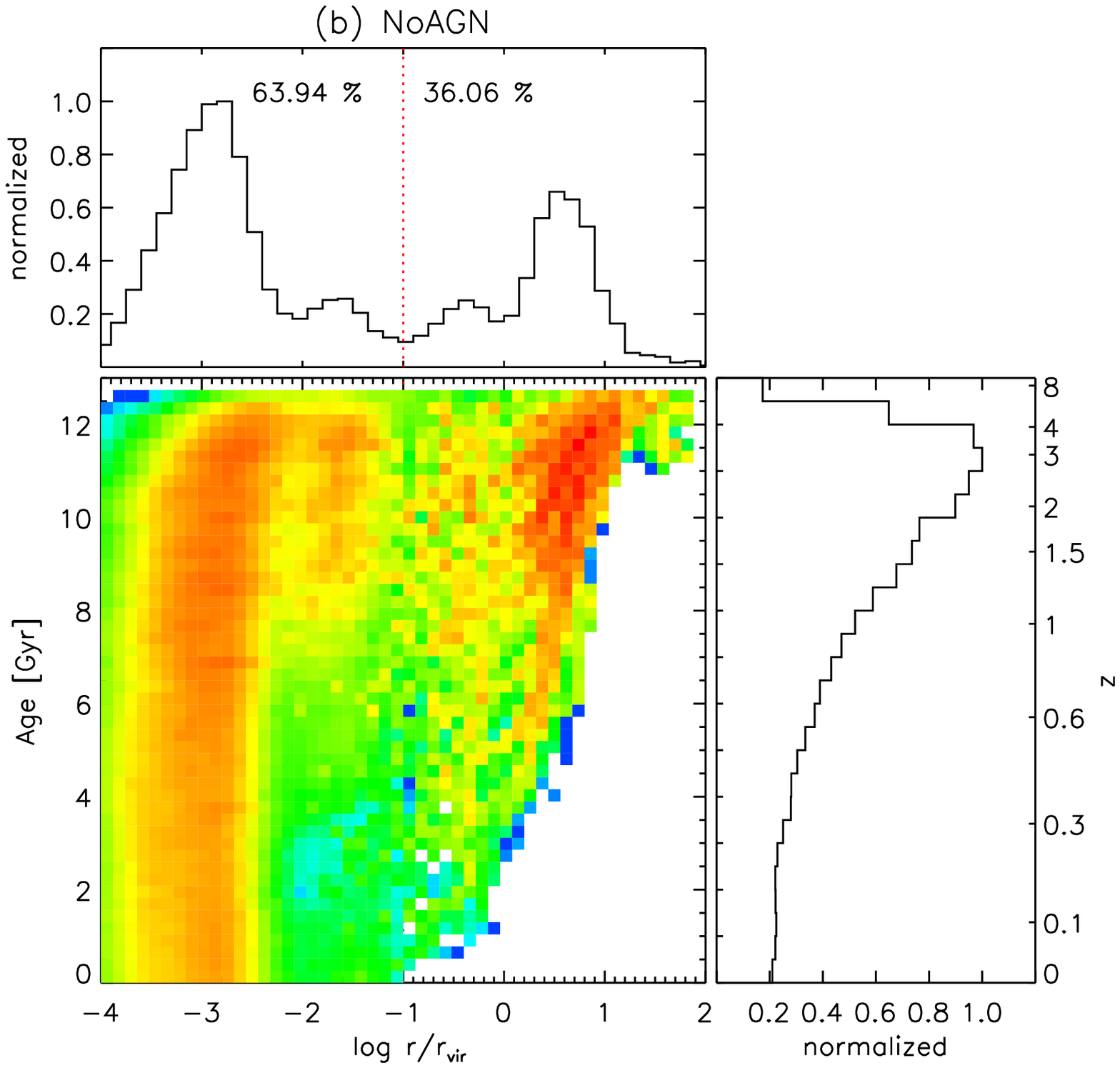}\\
    \includegraphics[width=.4\textwidth]{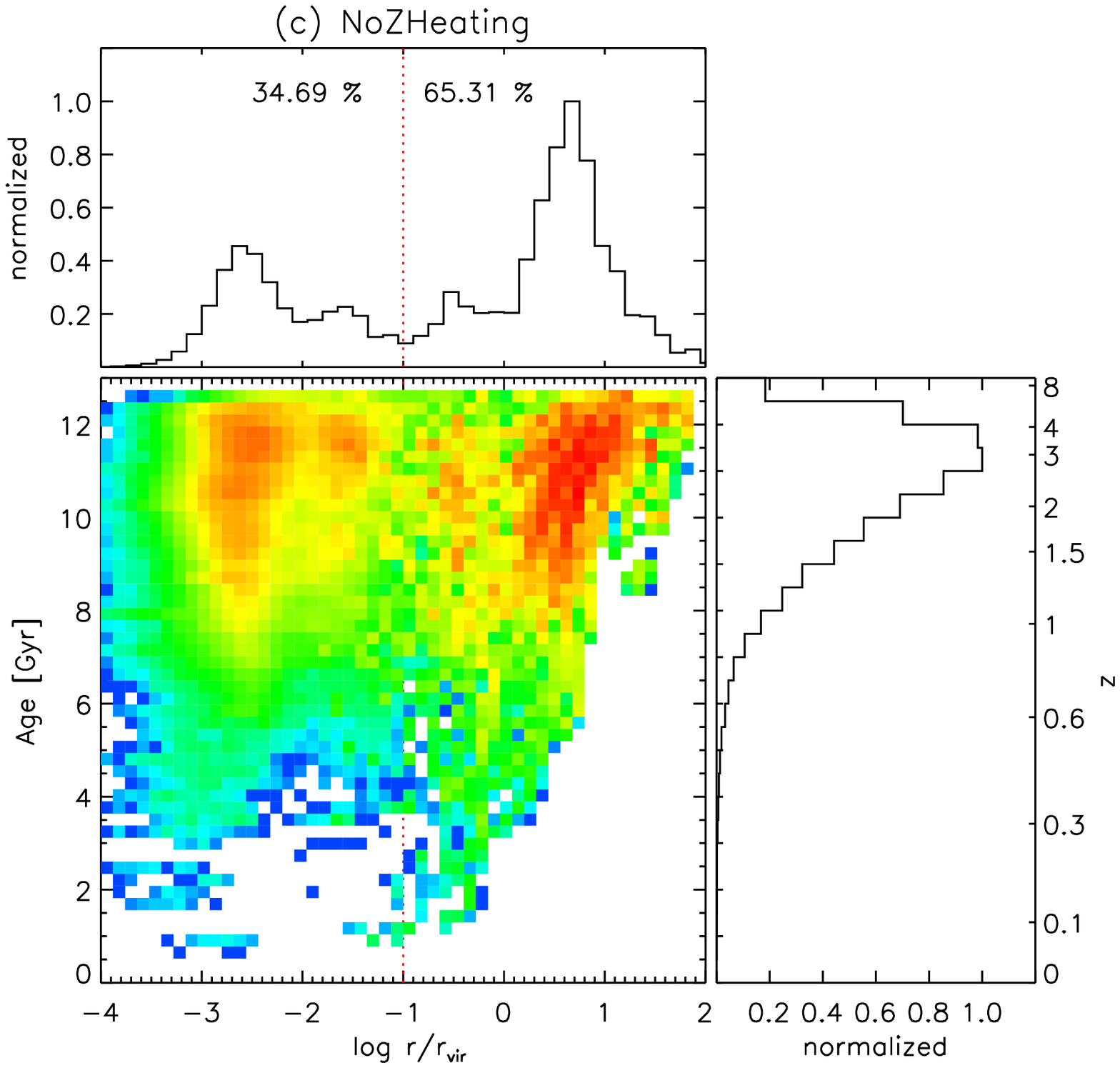}\quad
    \includegraphics[width=.4\textwidth]{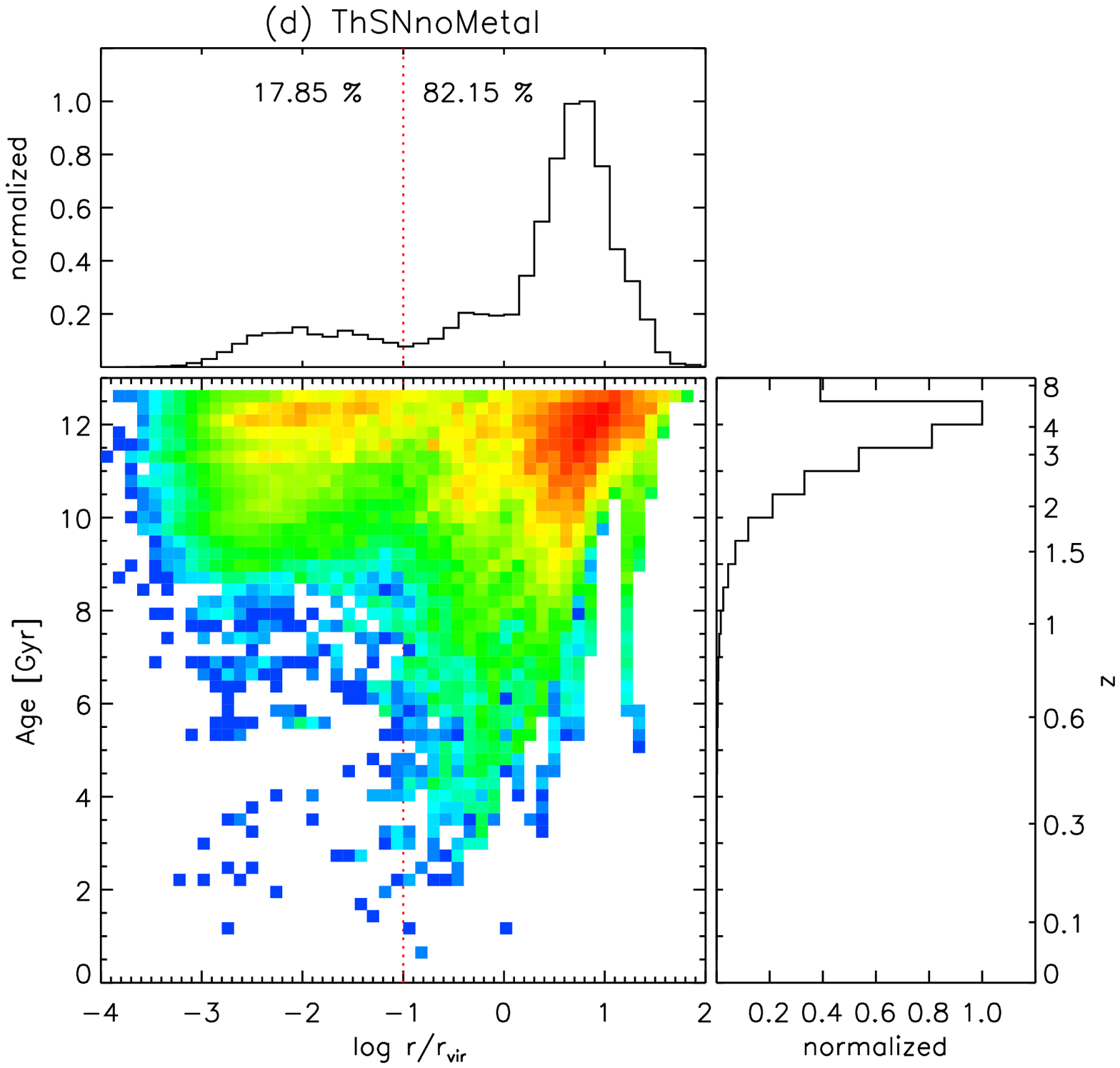}
    \caption{Stellar origin diagram for all stars within $r_{10}$ of 30 galaxies at z=0 
    in each model (a) Fiducial, (b) NoAGN, (c) NoZHeating and (d) ThSNnoMetal 
    model. Upper panels show the histogram of formation radii of stars and the 
    right panels show the star formation histories. The vertical red dotted line 
    shows the 10\% of virial radius, $r_{10}$.}     \label{fig:origin_all}
  \end{minipage}\\[1em]
\end{figure*}

\begin{figure*}
  \begin{minipage}{\textwidth}
    \centering
    \includegraphics[width=.3\textwidth]{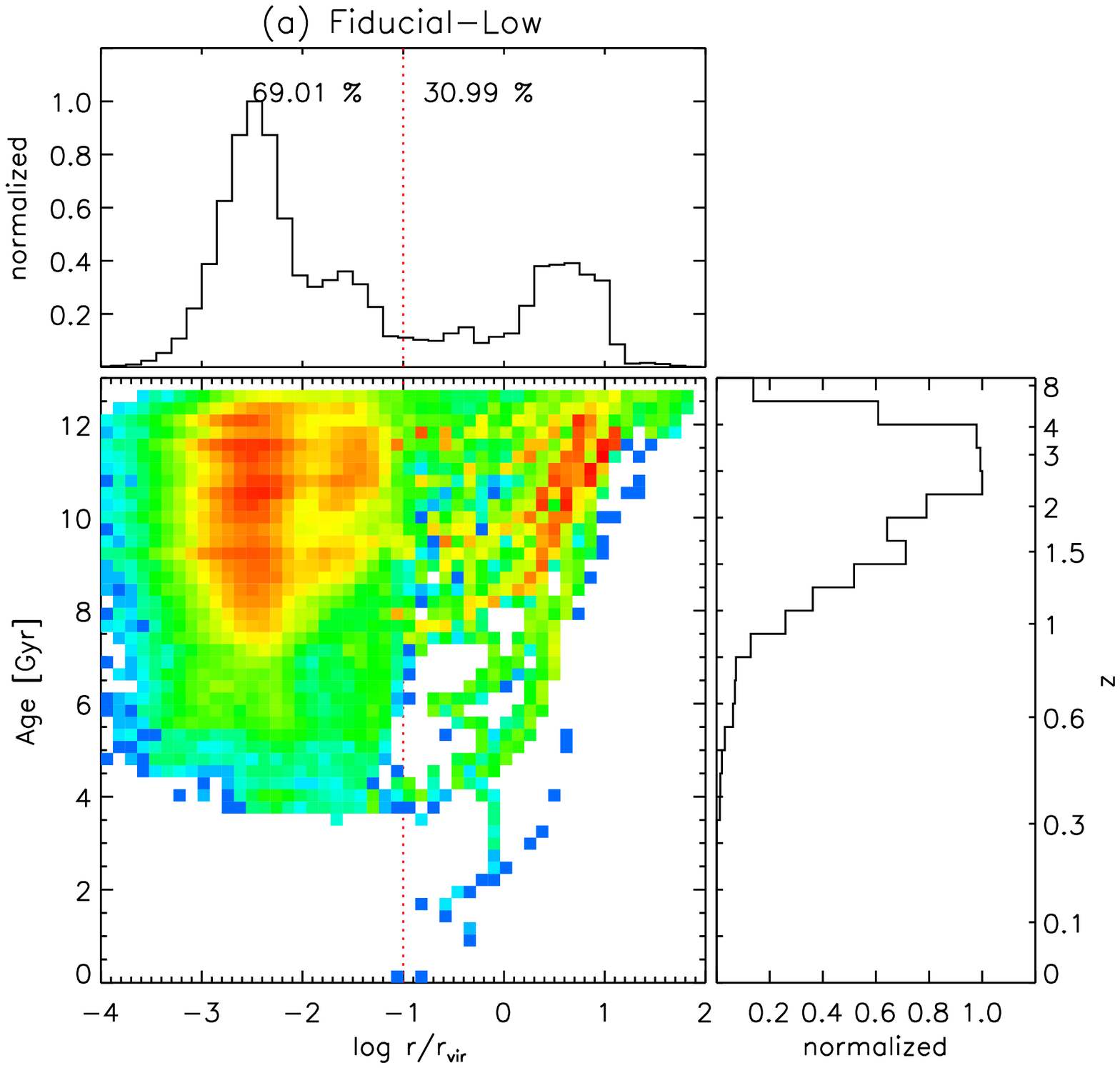}\quad
    \includegraphics[width=.3\textwidth]{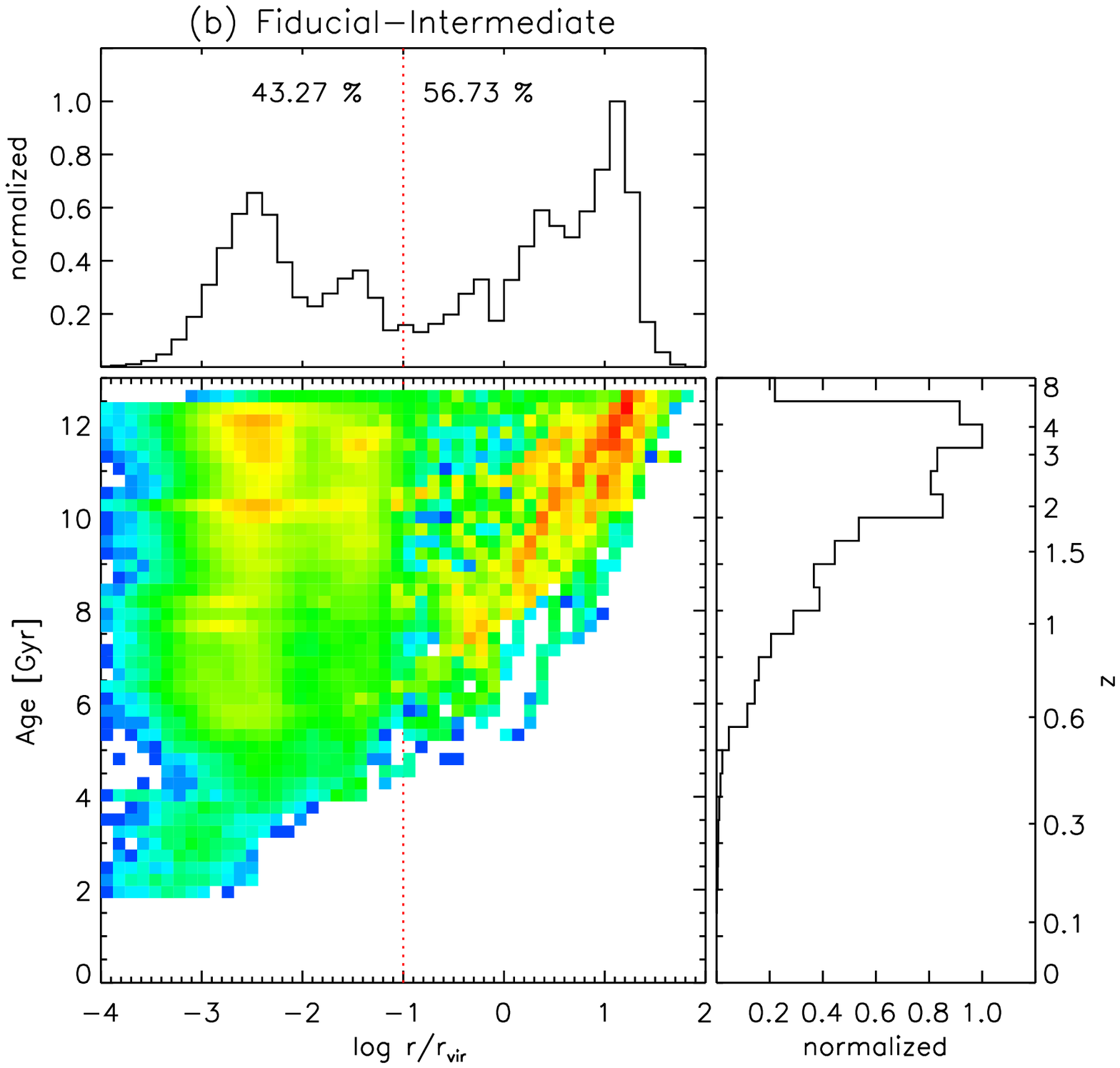}\quad
    \includegraphics[width=.3\textwidth]{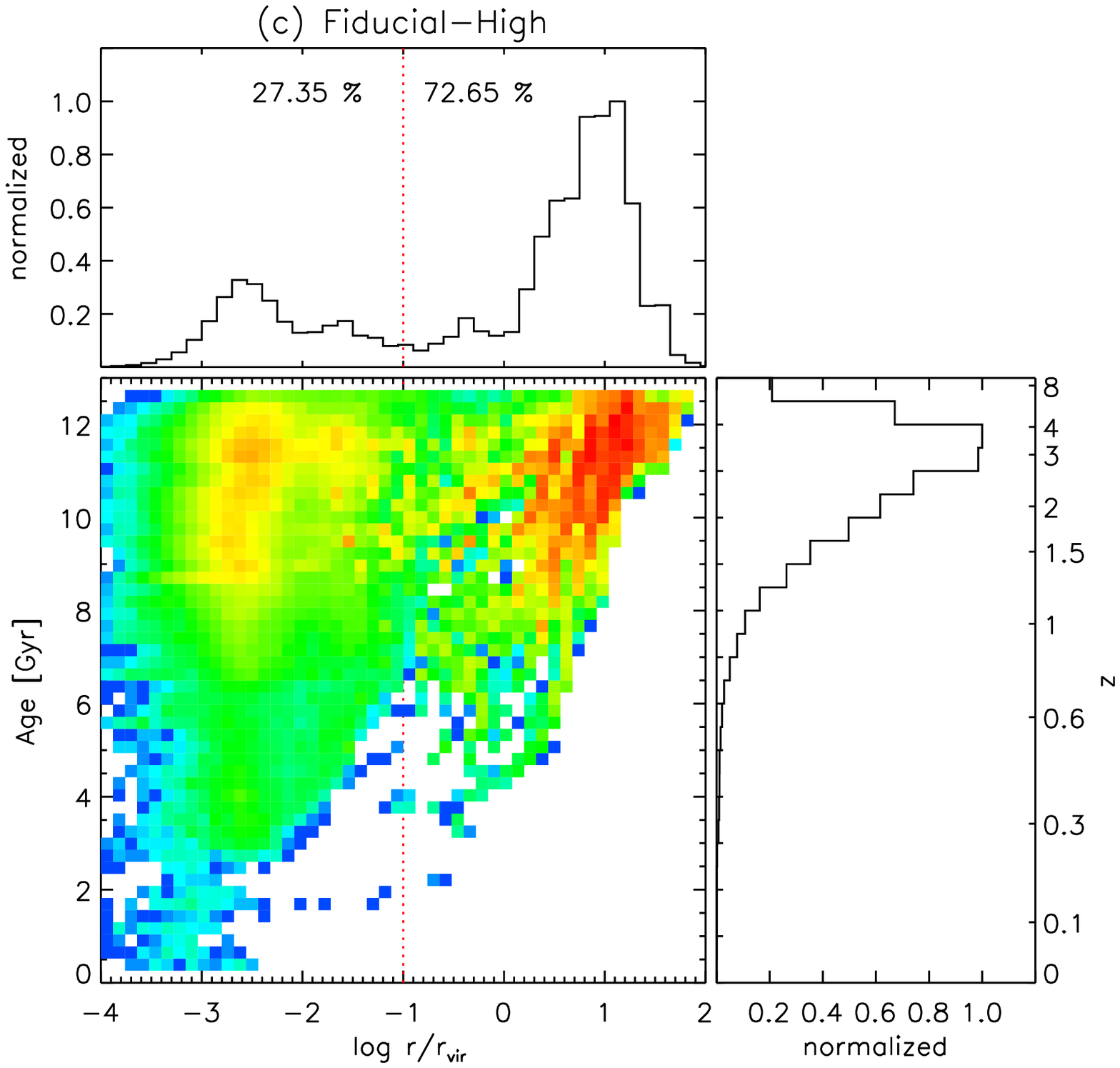}
    \caption{Stellar origin diagram for (a) low-mass ($8.5 \times 10^{10} < 
    M_{\ast}/\Msun < 1.25 \times 10^{11}$), (b) intermediate mass 
    ($1.25 \times 10^{11} < M_{\ast}/\Msun < 2.1 \times 10^{11}$), and (c) high 
    mass ($2.1 \times 10^{11} < M_{\ast}/\Msun < 5 \times 10^{11}$) galaxies from 
    the fiducial model. The ratio of in situ and ex situ stars are found to be sensitive 
    to the galaxy mass.}     \label{fig:origin_mass}
  \end{minipage}\\[1em]
\end{figure*}

Recent observations have established that early-type galaxies form and become 
red and dead early but continue to grow in mass and size without much late star 
formation \cite[e.g.][]{2005ApJ...626..680D,2006ApJ...650...18T,
2008ApJ...687L..61B,2012ApJ...749..121S}. These developments challenge 
classical formation models of early-type galaxies via monolithic collapse 
\citep{1962ApJ...136..748E} and equal-mass major mergers 
\citep[e.g.][]{1972ApJ...178..623T}, but favor a two-phase formation scenario 
\citep{2007ApJ...658..710N,2010ApJ...725.2312O}. In this scenario, the core of 
elliptical galaxies form early at $2<z<6$ by dissipational processes and cold gas 
flows \citep[e.g.][]{2009Natur.457..451D} and by merging of smaller structures of 
stars and gas. Then, the outer part of elliptical galaxy grows by accretion of old 
stars and galaxy mergers of all mass ratios \citep{2009ApJ...697.1290B,
2010ApJ...709.1018V,2012MNRAS.423.1544S,2012MNRAS.419.3200H}. 
But this subsequent build-up of the stellar envelop is dominated by 
non-dissipational processes via dry ``minor'' mergers as major mergers are rare 
at late times. This two-phase formation scenario can explain the  galaxy size 
growth and slight decline in velocity dispersion \citep{2012ApJ...744...63O}. The 
outer parts of ellipticals are found to be old and metal poor 
\citep{2015ApJ...807...11G} as would be expected if they had accreted at late 
times from low mass dwarf companions \citep{2012MNRAS.425..641L,
2015MNRAS.449..528H}.

We investigate the effect of various physics implemented in our simulation on the 
fundamental formation and assembly of galaxies. Figure~\ref{fig:origin_all} 
visualizes when and at which radius a star was born for star particles ending up 
within 10\% of virial radius of a present day galaxy. We have stacked all 30 
simulated galaxies for each physics model: (1) Fiducial (mechanical AGN and 
mechanical SN feebdack model), (2) NoAGN (without AGN feedback) (3) 
NoZHeating (without metal induced heating effect) and (4) ThSNnoMetal (with 
Thermal SN feedback and without metal enrichment). For stars that resident in 
final galaxies, some are made in situ, within the $r_{10}$ while some are made ex 
situ outside of the $r_{10}$ and later accreted. The vertical red dotted line 
indicates the 10\% of virial radius, $r_{10}$, which clearly separates these two 
phases as shown in the histogram of the formation radii in the upper panels of 
Figure~\ref{fig:origin_all}. We found two peaks at relatively similar location in the 
histogram of all four physics models: for the in situ formed stars at 
${\rm log}(r/r_{\mathrm vir}) \sim -2.5$ and for the ex situ formed stars at 
${\rm log}(r/r_{\mathrm vir}) \sim 0.7$ respectively. 

However, the ratio of in situ to ex situ stars is found to strongly vary with the 
feedback model we include. Our fiducial model galaxies on average have 40\% of 
stars formed in situ (see panel (a)), but this fraction increases to 64\% when we 
exclude the AGN feedback powered by the supermassive black holes. In addition, 
in the NoAGN feedback model the formation radius peak of in situ star formation
is located further inside of galaxies (${\rm log}(r/r_{\mathrm vir}) \sim -3.0$) 
compared to the fiducial model. The stellar feedback solely is not strong enough 
to generate galactic outflows in high-mass galaxies, and cannot quench in situ 
star formation. The central accretion of ambient gas from stellar mass loss 
continuously triggers star formation \citep{1997ApJ...487L.105C} and this late 
time star formation contributes significantly to the in situ fraction of stars.

In the NoZHeating model, when we exclude the metal induced heating, the 
fraction of ex situ formed star increases from 60\% to 65\%. These extra heating 
effects (photoelectric and cosmic X-ray background heating) mainly affect the 
small systems which later accrete on to the central galaxies, preventing small 
blobs from condensing. This tends to decrease the accreted star fraction.

The ratio of in situ and ex situ stars also strongly depends on {\it how} the 
feedback prescription is implemented in the model. In the ThSNnoMetal model, 
where we use thermal SN feedback, which does not produce strong SN winds 
compared to the mechanical SN and early stellar feedback, the fraction of ex 
situ formed star increases to 82\%.  

We divide our simulated galaxies into three mass bins: high ($2.1 \times 10^{11} 
< M_{\ast}/\Msun < 5 \times 10^{11}$), intermediate ($1.25 \times 10^{11} 
< M_{\ast}/\Msun < 2.1 \times 10^{11}$) and low ($8.5 \times 10^{10} < 
M_{\ast}/\Msun < 1.25 \times 10^{11}$) so that we have 10 galaxies in each mass 
bin. The ratio of in situ and ex situ stars is found to be sensitive to the galaxy 
mass as found in \cite{2010ApJ...725.2312O,2017MNRAS.464.1659Q,
2016MNRAS.458.2371R}, i.e., more stars form in situ for low mass galaxies and 
more in ex situ for higher mass galaxies. Also, there is a continuous trend of in 
situ star formation quenching redshift with stellar mass. While low mass galaxies 
show high in situ star formation rate until redshift $z=1$, high mass galaxies 
quench early before $z=1$. Some of our high mass galaxies are rejuvenated 
shortly due to the recycled gas ejected from  the old stellar population within the 
innermost region of the galaxies (log $r_{10} / r_{\mathrm vir} \sim -3$).

\section{Summary and discussion}\label{sec:dis}
In this work, we present a more comprehensive treatment of the physical 
processes consequent to the allowance of chemical evolution highlighting the 
extra heating processes due to the interaction of metals and radiation. We then 
used updated simulations of 30 massive galaxies with halo mass of 
$\Mvir \sim 10^{12-13} \Msun$ at $z=0$ to make comparisons between the 
simulation and observations. The feedback models account for (1) AGN feedback 
via broad absorption line winds and X-ray radiation heating, (2) stellar feedback 
via UV heating and stellar winds from young and old stars, and supernovae 
type~I and II as well as associated mass and metal generation, (3) additional 
metal heating effect via photoelectric heating and cosmic X-ray background 
heating from accreting black holes in background galaxies. Overall energy and 
momentum budget from all feedback effects has been shown to generate realistic 
galaxy properties (fiducial model). We also explore the detailed role of separate 
feedback variables by running three additional sets of simulations without AGN 
feedback (NoAGN model), without metal heating effects (NoZHeating model) and 
without the suite of mechanical stellar feedback but with thermal SN feedback 
(ThSNnoMetal model) to better understand the separate effects of various 
physical processes. The input physics variation of the simulation suite is 
summarized in Table~\ref{tab:models}.

We find that the AGN feedback plays the most dominant role in reproducing the 
basic physical properties of observed massive early type galaxy. The mechanical 
and radiative AGN feedback generates strong, large scale outflows during early 
stages of galaxy evolution and efficiently quenches in situ star formation in 
massive galaxies. In the absence of AGN feedback, the conversion efficiency of 
baryons into stars in the central galaxies is increased by a factor of 3 
(Figure~\ref{fig:fbar}), and all simulated galaxies never quench and keep forming 
stars throughout to $z=0$ (Figure~\ref{fig:sfr}). The formation of central galaxies 
is dominated by in situ star formation without AGN feedback 
(Figure~\ref{fig:origin_all}(b)), and this leads to very compact stellar cores with a 
factor of five smaller effective radii than in the observed mass-size relation and 
high velocity dispersions (Figure~\ref{fig:size},\ref{fig:sigma}). AGN feedback can 
efficiently drive out and remove gas from the galaxy halos and prevent large gas 
concentrations in halos, resulting in much lower X-ray luminosities and gas mass 
fractions (Figure~\ref{fig:lx},\ref{fig:fgas}).

We show that the metal induced heating mechanisms primarily affect diffuse and 
smaller galaxies at high redshift. They prevent baryons from condensing and 
accumulating in dense small blobs mainly at high redshift and therefore delay star 
formation. Star formation, however, later compensates, as the heated gas cools
and comes back to the central galaxies. Therefore, metal induced heating 
mechanisms have a negligible effect on the final stellar mass as well as the final 
gas properties of central galaxies. However, they certainly reduce the mass of 
small stellar systems later accreted to the central galaxy. The fraction of accreted 
stars is reduced from 65\% to 60\% when we add metal induced heating 
(Figure~\ref{fig:origin_all}(c)), and this results in slightly smaller stellar galaxy
sizes as they tend to have fewer number of dissipationless low-redshift mergers. 
Although the metal induced heating overall produces weak effects on these 
stellar mass scale ($M_{\ast} \sim M^{11-12} \Msun$), they can have a dramatic
impact on the evolution of dwarf galaxies with much smaller masses. For 
example, \citet{2016Natur.535..523F} recently showed that the photoelectric 
heating plays a more important role than SN feedback in regulating star 
formation in dwarf galaxies with stellar mass of $10^7 \Msun$.

Although  AGN feedback seems to exert the dominant impact on massive galaxy 
formation, it is also crucial to properly include the stellar feedback to regulate the 
star formation in small building blocks and to suppress in situ star formation at 
high redshift \citep{2013MNRAS.436.2929H}. We show that inclusion of thermal 
SN feedback instead of ejective SN feedback yields final massive galaxies 
dominated by hierarchically assembled small galaxies, therefore much higher 
accreted fraction (Figure~\ref{fig:origin_all}(d)).

Overall, we achieve fairly good agreement between our fiducial model and 
observational constraints in the local Universe, although some small 
discrepancies still remain. To some extent, this may be linked to further physical 
mechanisms which have not been included so far such as cosmic rays, 
magnetohydrodynamics, and effects of relativistic jets from AGN. In particular, 
the simulations do not include cosmic ray driven winds 
\citep[e.g.][]{2013ApJ...777L..16B,2013ApJ...777L..38H,2014MNRAS.437.3312S,
2016arXiv160802585W} and the effect of runaway OB stars that migrate into 
low-density regions \citep{2015ApJ...814....4L} and their absence may contribute 
to the excess of stellar masses in massive galaxies. In future work, we will extend 
our study to include several of these effects all of which can act directly on the 
gas and reduce the star formation rate.

\begin{acknowledgements}
We are grateful to the anonymous referee for very helpful comments 
on the manuscript. We also thank Volker Springel for making  the GADGET-3 
code available and Sergey Sazonov for helpful discussions. Support for this 
work is provided by NASA through grant number HST Cycle 23 AR-14287 
from the Space Telescope Science Institute. TN acknowledges support 
from the DFG cluster of excellence ``Origin and Structure of the Universe''. 
MH acknowledges financial support from the European Research Council 
via an Advanced Grant under grant agreement no. 321323 NEOGAL. 
Numerical simulations were run on the compute facility of the Princeton 
Institute of Computational Science and engineering. 
\end{acknowledgements}

\bibliography{references}
\appendix\label{sec:appendix}
\begin{figure}[b]
\epsfig{file=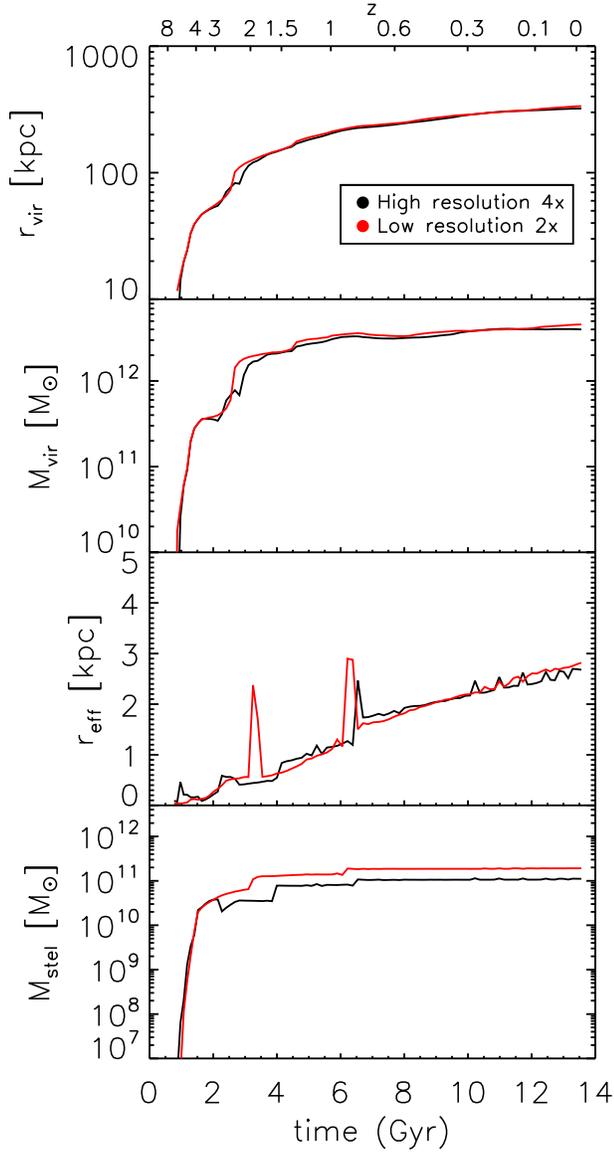,width=\columnwidth}
\caption{The cosmic evolution of  virial radius, virial mass, effective radius and 
stellar mass of the fiducial run of halo m0290 at two different resolutions (black: 
high resolution, red: fiducial resolution). We find no significant difference in the 
physical properties of central galaxy with different resolution.}
\label{fig:convergence}
\end{figure}

\section{Convergence with Resolution} 
We test the convergence of the physical properties of central galaxy with 
respect to resolution using two realization of the halo m0290 with 
$M_{\rm vir} \sim 4 \times 10^{12} \Msun$. In Figure~\ref{fig:convergence} 
we label the fiducial m0290 simulation `fiducial resolution' and ran a higher 
resolution simulation. The `high resolution' run has a gas and dark matter 
particle mass eight times smaller and a gas softening length twice smaller 
than the reference resolution. The high resolution run is simulated with the 
number of supernova events reduced by a factor of three but with increased 
SN feedback energy in order to account for clumping of star formation but 
otherwise identical parameters. The agreement between the high resolution 
and low resolution runs is very good.

\end{document}